\documentclass[journal]{IEEEtran}
\usepackage{amsmath,amsfonts}
\usepackage{algorithmic}
\usepackage{algorithm}
\usepackage{gensymb}
\usepackage{array}
\usepackage[caption=false,font=normalsize,labelfont=sf,textfont=sf]{subfig}
\usepackage{textcomp}
\usepackage{stfloats}
\usepackage{url}
\usepackage{verbatim}
\usepackage{graphicx}
\usepackage{diagbox}
\usepackage{cite}
\begin{document}

%
%
\title{Deep Reparameterization for Full Waveform Inversion: Architecture Benchmarking, Robust Inversion, and Multiphysics Extension}

\author{Feng Liu, Yaxing Li, Rui Su, Jianping Huang, Lei Bai
\thanks{Manuscript submitted April 24, 2025}
\thanks{This research is supported by the the Shanghai Municipal Science and Technology Major Project (\textit{Corresponding author: Rui Su}).}
\thanks{Feng Liu is with the School of Electronic Information and Electrical Engineering, Shanghai Jiao Tong University, Shanghai 200240, China, and also with Shanghai Artificial Intelligence Laboratory, Shanghai 200232, China (e-mail: liufeng2317@sjtu.edu.cn). This work was done during his internship at Shanghai Artificial Intelligence Laboratory.}
\thanks{Yaxing Li and Jianping Huang are with the Chengdu University of Technology, Chengdu 610059, China (email: yxli2024@cdut.edu.cn; jphuang@upc.edu.cn).}
\thanks{Rui Su and Lei Bai are with the Shanghai Artificial Intelligence Laboratory, Shanghai 200232, China (email: surui@pjlab.org.cn; bailei@pjlab.org.cn).}
}

\markboth{
}%
{Shell \MakeLowercase{\textit{et al.}}: A Sample Article Using IEEEtran.cls for IEEE Journals}



\maketitle

%
%
\begin{abstract}
    Full waveform inversion (FWI) is a high-resolution subsurface imaging technique, but its effectiveness is limited by challenges such as noise contamination, sparse acquisition, and artifacts from multiparameter coupling. To address these limitations, this study develops a deep reparameterized FWI (DR-FWI) framework, in which subsurface parameters are represented by a deep neural network. Instead of directly optimizing the parameters, DR-FWI optimizes the network weights to reconstruct them, thereby embedding structural priors and facilitating optimization. To provide benchmark guidelines for the design of DR-FWI, we conduct a comparative analysis of three representative architectures (U-Net, CNN, MLP) combined with two initial model embedding strategies: one pretraining the network to generate predefined initial models (pretraining-based), while the other directly adds network outputs to the initial models. Extensive ablation experiments show that combining CNN with pretraining-based initialization significantly enhances inversion accuracy, offering valuable insights into network design. To further understand the mechanism of DR-FWI, spectral bias analysis reveals that the network first captures low-frequency features and gradually reconstructs high-frequency details, enabling an adaptive multi-scale inversion strategy. Notably, the robustness of DR-FWI is validated under various noise levels and sparse acquisition scenarios, where its strong performance with limited shots and receivers demonstrates reduced reliance on dense observational data. Additionally, a “backbone-branch” structure is proposed to extend DR-FWI to multiparameter inversion, and its efficacy in mitigating cross-parameter interference is validated on a synthetic anomaly model and the Marmousi2 model. These results suggest a promising direction for joint inversion involving multiple parameters or multiphysics.
\end{abstract}


%
%
\section{INTRODUCTION}
    \IEEEPARstart{F}{ull} Waveform Inversion (FWI) is a wave-equation-based high-resolution subsurface parameter estimation method that minimizes the residuals between simulated and observed seismic wavefields through iterative optimization \cite{lailly_1983_Seismic,tarantola_1984_Linearized}. Compared to conventional approaches such as ray-based tomography and travel-time imaging, FWI demonstrates superior capabilities by synergistically utilizing full waveform features including amplitude, phase, and travel-time information \cite{virieux_2009_Overview, tromp_2019_Seismic}. This comprehensive data exploitation enables exceptional performance in complex geological structure inversion (e.g., salt domes \cite{kalita_2019_Regularized}, fault systems \cite{brossier_2009_Seismic}) and resource exploration applications (particularly in hydrocarbon reservoirs \cite{routh_2017_Impact} and geothermal fields \cite{schmelzbach_2016_Advanced}). The method's essence lies in its fundamental departure from the geometric approximations inherent in ray theory, instead employing numerical solutions of the wave equation to achieve high-fidelity parameter reconstruction. This capability provides critical petrophysical constraints for engineering decision-making process in exploration geophysics \cite{virieux_2009_Overview}.

    Despite its theoretical potential, FWI faces substantial challenges in practical implementations. First, the highly nonlinear nature of the objective function makes the inversion process susceptible to entrapment in local minima \cite{gauthier_1986_Twodimensional, luo_1991_Waveequation}. Gradient-based algorithms often exhibit poor convergence robustness when the initial model substantially deviates from the true subsurface properties \cite{métivier_2014_Full}. Second, the ill-posedness of the inverse problem manifests as solution non-uniqueness, where distinct parameter combinations can produce nearly identical wavefield responses, thereby amplifying solution uncertainty \cite{tarantola_2005_Inverse}. Moreover, in multiparameter joint inversions (e.g., P-wave velocity ($v_p$) , S-wave velocity ($v_s$), and density ($\rho$)), crosstalk effects between parameters can introduce spurious anomalies, critically undermining the physical reliability of inverted results \cite{operto_2013_Guided, yang_2014_Multiparameter}. These intertwined issues collectively dictate that FWI's practical efficacy remains heavily contingent upon initial model accuracy, observational system completeness, and appropriate regularization strategy design.

\begin{figure*}[!ht]
    \centering
    \includegraphics[width=1.0\textwidth]{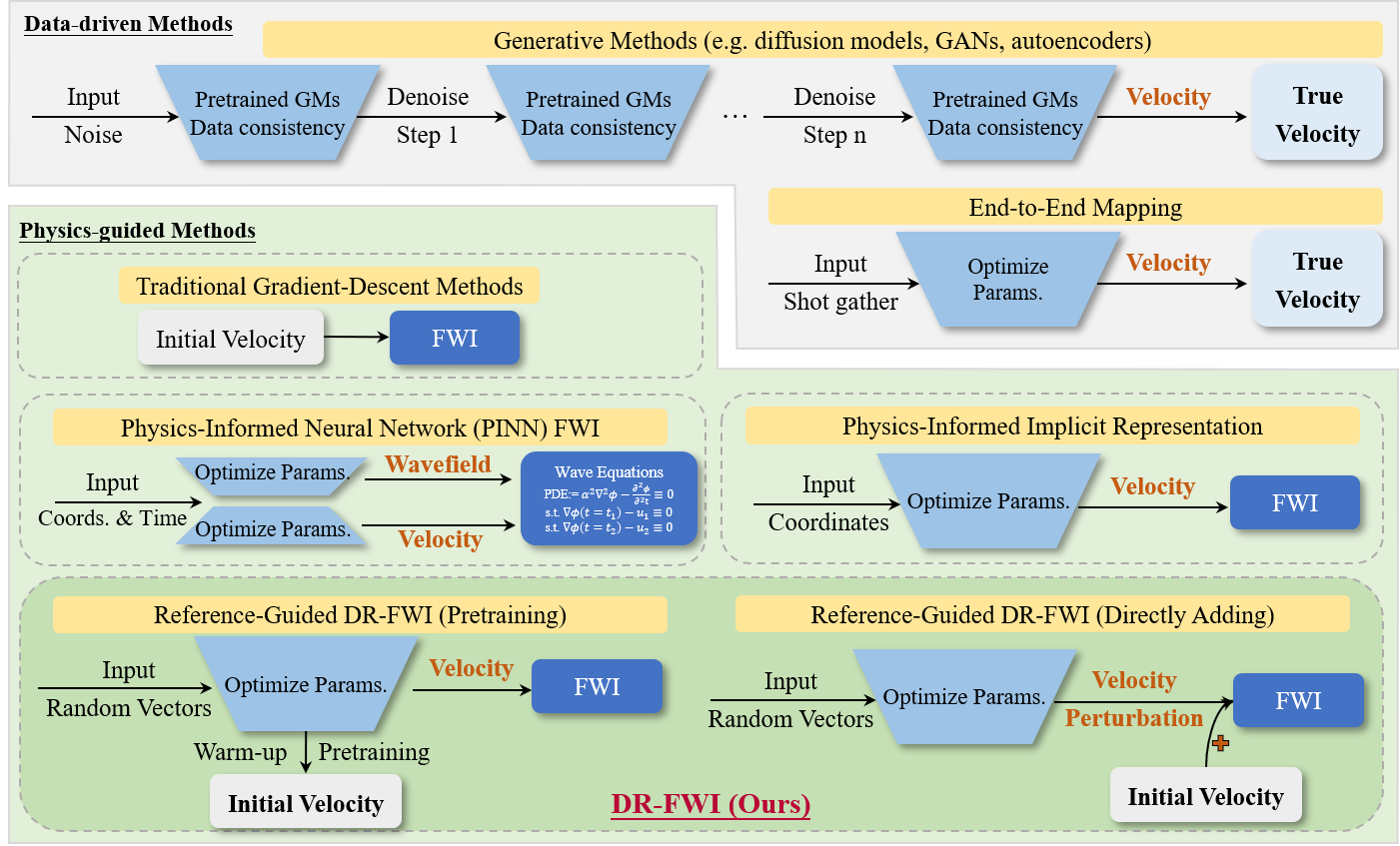}
    \caption{Comparison of data-driven methods (e.g., generative model-based, end-to-end supervised learning) and physics-driven methods (e.g., traditional FWI, PINNs-based FWI, implicit representation-based IFWI), along with the two forms of reference-guided DR-FWI proposed in this study: one pretraining the network to generate predefined initial models, while the other directly adds network outputs to the initial model.}
    \label{fig:fwi_types}
\end{figure*}

    Recent advances in deep learning have provided new pathways to overcome traditional FWI limitations \cite{yu_2021_Deep,lin_2023_Physicsguided}. Current methodologies for subsurface inversion can be broadly categorized into two paradigms: data-driven approaches employ end-to-end networks to directly map observed data to subsurface medium parameters \cite{wu_2020_InversionNet, zhang_2020_Datadriven}, yet often suffer from a strong dependency on annotated datasets and poor physical interpretability; physics-driven methods embed wave-equation constraints into hybrid inversion frameworks \cite{rasht‐behesht_2022_Physicsinformed,ren_2020_Physicsbased}, offering greater resilience under sparse data conditions and improved consistency with physical laws. As an example, Fig.~\ref{fig:fwi_types} illustrates a subset of research on data-driven and physics-driven FWI, including approaches based on generative models and end-to-end mapping for the former, and physics-informed neural networks (PINNs) as well as deep re-parameterization techniques for the latter. As a pivotal branch of physics-driven paradigm, deep reparameterization reformulates inversion parameters as low-dimensional latent variables via neural network-based implicit representations. This reformulation not only mitigates the nonlinearity of objective function but also introduces learnable regularization into the inversion process \cite{he_2021_Reparameterized,zhu_2021_Integrating,sun_2023_Implicit,liu_2025_Automatic}. Prior studies have successfully applied this strategy to construct prior constraints, reduce dependence on initial models, and mitigate multiparameter crosstalk. To enhance regularization, Wu \& McMechan (2019)\cite{wu_2019_Parametric} and He \& Wang (2021)\cite{he_2021_Reparameterized} implemented a multi-layer convolutional neural networks (CNNs) for velocity models parameterization, employing a two-phase training scheme: Phase I pretrains the network to map random vectors to initial velocity models, while Phase II integrates adjoint-state methods for physics-constrained optimization. Their synthetic experiments demonstrated improved peak signal-to-noise ratio compared to conventional FWI. Zhu et al. (2021)\cite{zhu_2021_Integrating} adopted shallower CNNs to generate velocity perturbations, which were superimposed on background models as inputs to the FWI process. This architecture exhibited enhanced noise robustness when tested on data contaminated with Gaussian noise. For addressing initial model dependency, Sun et al. (2023)\cite{sun_2023_Implicit} proposed an implicit FWI framework using fully-connected networks (FNNs) or multi-layer perceptrons (MLPs) to directly map spatial coordinates to subsurface velocity values, thereby removing the reliance on initial velocity models. Notably, Sun et al. (2021)\cite{sun_2021_Physicsguided} pioneered a discriminative autoencoder for data-physics co-driven FWI, which was subsequently extended by Wang et al. (2023)\cite{wang_2023_Prior} and Taufik et al.(2024) \cite{taufik_2024_Learned} through diffusion model-based generative architectures. While these hybrid methods alleviate the dependence on initial velocities, they still require massive paired velocity-seismic datasets for effective network training. In multiparameter inversion, Dhara et al. (2022)\cite{dhara_2022_ElasticAdjointNet} developed an autoencoder-based collaborative framework, in which an encoder compresses multi-shot seismic records into latent representations, and parallel decoders generate distinct subsurface parameter perturbations for background models superposition. 
    
    Despite recent progress, several critical limitations hinder the broader applicability of deep reparameterization-based FWI. First, the lack of theoretical guidance on networks design and hyperparameters tuning often leads to heuristic architectures with limited transferability. Second, most existing studies focus predominantly on noise robustness, with insufficient investigation into other practical scenarios such as sparse acquisitions. Third, multiparameter inversion remains underexplored; existing approaches rely on case-specific designs for limited parameter sets, lacking a unified and scalable framework applicable to diverse physical properties. Moreover, the absence of open-source implementations restricts reproducibility and further development.

    To address these challenges, this paper proposes a reference-guided deep reparameterization FWI framework (DR-FWI). The core innovation lies in the integration of prior geological knowledge via pretraining paradigms and the design of multiparameter adaptive architectures. The main contributions of this work are summarized as follows:
    \begin{enumerate}
        \item \textbf{Comprehensive Evaluation of Network Architectures and Reference Integration Strategies}: We systematically evaluate a range of network architectures (CNN, MLP, U-Net) along with strategies for incorporating reference velocity models, including stepwise pretraining-based embedding and direct model superposition. Experimental results demonstrate that shallow CNNs effectively capture complex geological structures and outperform both the conventional FWI and deeper network counterparts. Moreover, the stepwise embedding approach consistently delivers superior inversion accuracy and convergence compared to direct superposition, providing practical guidance for the design of network architectures and reference-guided embedding strategies in field applications.
        \item \textbf{Improved Robustness under Data-Scarce and Noisy Conditions}: We provide the evidence that DR-FWI exhibits highly resilient under extremely sparse acquisitions (e.g., 10 sources × 20 receivers) and severe noise contamination (e.g., Gaussian noise with a sixfold standard deviation). Compared to the conventional FWI, the proposed approach substantially reduces inversion errors while significantly relaxing data acquisition requirements, thereby extending the applicability of FWI to underexplored regions and cost-sensitive scenarios.
        \item \textbf{Adaptive multiparameter Joint Inversion Architecture}: Building upon single-parameter reparameterization CNNs, we design a unified backbone–branch architecture for multiparameter FWI. The shared backbone captures common structural features across different physical parameters, while separately branches extract parameter-specific characteristics. Each branch output is further scaled to its respective physical range via parameter-wise normalization and de-normalization operations. By applying this architecture to the joint inversion of $v_p$, $v_s$ and $\rho$ in both anomaly models and the Marmousi2 benchmark, numerical experiments demonstrate its effectiveness in substantially mitigating crosstalk interference. This framework establishes a scalable and accurate solution for multi-physics joint inversion, enabling more reliable reconstruction of complex subsurface properties.
    \end{enumerate}
    This study advances physics-constrained deep learning inversion along three key dimensions: architectural benchmarking, scenario generalization, and multi-physics extensibility. Experimental results confirm the seamless integration of deep reparameterization into conventional FWI workflows, establishing it a versatile framework for resource exploration and geohazard early-warning systems. Furthermore, we explain how the low-frequency bias in deep neural networks contributes to the observed improvements in FWI, with the deep reparameterization network's ability to capture low-frequency components enhancing both inversion accuracy and robustness. These findings prove the practical effectiveness of the proposed method in challenging geophysical inversion tasks.

%
%

\section{Methodology}

\subsection{Physics-Guided FWI}

    Traditional FWI is entirely physics-driven inversion approach that employs gradient-based optimization methods to iteratively minimize the mismatch between observed and simulated seismic data. The physical principles underlying the inversion are governed by wave equations, which establish the relationship between subsurface properties and recorded waveforms. In a two-dimensional (2-D) acoustic medium, the governing equations in time domain are given by \cite{alford_1974_ACCURACY, schuster_2017_Seismic}:
    \begin{align}\label{eqn:acoustic}
    \left\{
    \begin{aligned}
        \frac{\partial v_i}{\partial t} &= \frac{1}{\rho} \frac{\partial p}{\partial x_i} \\
        \frac{\partial p}{\partial t} &= \kappa \nabla \cdot \textbf{v} + f
    \end{aligned},
    \right.
    \end{align}
    where $\rho$ denotes the density, $v_i$ represents the particle velocity in the $i$-th direction, $p$ denotes the pressure, $f$ is the source term, and $\kappa$ is the bulk modulus. 
    
    To account for more complex and realistic subsurface characteristics, the elastic wave equation, which incorporates multiple elastic moduli, can be written as \cite{levander_1988_Fourthorder,virieux_1986_PSV}:
    \begin{align}\label{eqn:elastic}
    \left\{
    \begin{aligned}
        \rho \frac{\partial v_i}{\partial t} &= \frac{\partial \tau_{ij}}{\partial x_j} + f_i \\
        \tau_{ij} &= C_{ijkl}\epsilon_{kl}
    \end{aligned},
    \right.
    \end{align}
    where $\tau_{ij}$ and $\epsilon_{kl}$ represent the stress and strain tensors, respectively, $f_i$ is the external body force in the $i$-th direction, and $C_{ijkl}$ is the elastic moduli tensor, with indices $i, j, k, l \in { x, y, z}$. A detailed implementation of different wave equation formulations can be found at Liu et al. (2025) \cite{liu_2025_Automatic}. All subsequent experiments in this study are conducted within the flexible ADFWI framework \cite{liu_2024_Liufeng2317}.

    The general form of equations~\ref{eqn:acoustic} and \ref{eqn:elastic} can be expressed as: 
    \begin{align}\label{eqn:wave_equation}
        \textbf{d} = F(\textbf{m}),
    \end{align}
    where $F$ represents the forward operator, $\textbf{d}$ denotes the synthetic seismic data, and $\textbf{m}$ describes the subsurface properties. Specifically, in an acoustic media, $\textbf{m}$ typically consists of the P-wave velocity ($v_p$) and density ($\rho$); in isotropic elastic media, $\textbf{m}$ is extended to include the P-wave velocity, S-wave velocity ($v_s$), and density; and in anisotropic elastic media, $\textbf{m}$ further incorporates anisotropic parameters such as $\epsilon$, $\delta$, and $\gamma$.

    The classical FWI with a general regularization term can be formulated as a constrained optimization problem:
    \begin{align}\label{eqn:objective_function}
        \textbf{m}^* = \underset{\textbf{m}}{\arg\min} \{ \mathcal{D}(d_{obs}, d_{cal}(\mathbf{m})) + \lambda \mathcal{R}(\textbf{m})\},
    \end{align}
    where $\textbf{m}^*$ represents the final inverted subsurface model, $\mathcal{D}$ measures the misfit between observed data $d_{obs}$ and synthetic data $d_{cal}(\textbf{m})$, $\mathcal{R}(\textbf{m})$ is a regularization term, and $\lambda$ is the corresponding weight. The misfit function can be defined using various norms, such as the L2-norm, the global cross-correlation, or alternative metrics such as the envelope misfit \cite{lailly_1983_Seismic, bozdağ_2011_Misfit, choi_2012_Application}.
    
    The commonly used optimization strategy in FWI is gradient descent-based optimization, which iteratively refines an initial subsurface model ($\mathbf{m}_0$) until the synthetic data sufficiently match the observed data. Specifically, at the $(k+1)$-th iteration, the model parameters are updated as:
    \begin{align}\label{eqn:traditional_optimization}
        \textbf{m}_{k+1} = \textbf{m}_k + \alpha_k \frac{\partial \mathcal{L}}{\partial \textbf{m}_k} 
    \end{align}
    where $\mathcal{L}$ denotes the objective function to be minimized, and $\partial \mathcal{L}/\partial \textbf{m}_k$ denotes the descent direction of the objective function, which can be computed using the adjoint-state method \cite{fichtner_2006_Adjoint, liu_2006_FiniteFrequency} or automatic differentiation \cite{sambridge_2007_Automatic,liu_2025_Automatic}, and $\alpha_k$ is the step length.

    In addition to traditional gradient descent-based methods, the recent surge in deep learning has spurred interest in physics-driven deep learning approaches for FWI. For example, physics-informed neural networks (PINNs) aim to learn mappings from spatial coordinates $(x, y, z)$ to both seismic wavefields and the subsurface properties using separate neural networks \cite{rasht‐behesht_2022_Physicsinformed, herrmann_2023_Use}. These networks incorporate physical laws by enforcing the wave equation constraints (e.g., equation~\ref{eqn:acoustic} or equation~\ref{eqn:elastic}) as soft penalties within the loss function, thereby embedding physics directly into the learning process. Although PINN-based methods have shown promising performance in synthetic experiments, they still face several critical challenges such as limited training data, high computational cost, and difficulties in generalizing to field data. These limitations highlight the need for an inversion strategy that more effectively integrates the strengths of deep neural networks and physical equations, improving the performance of conventional FWI without significantly increasing computational burden and data requirements.

\begin{figure*}[!ht]
    \centering
    \includegraphics[width=1.0\textwidth]{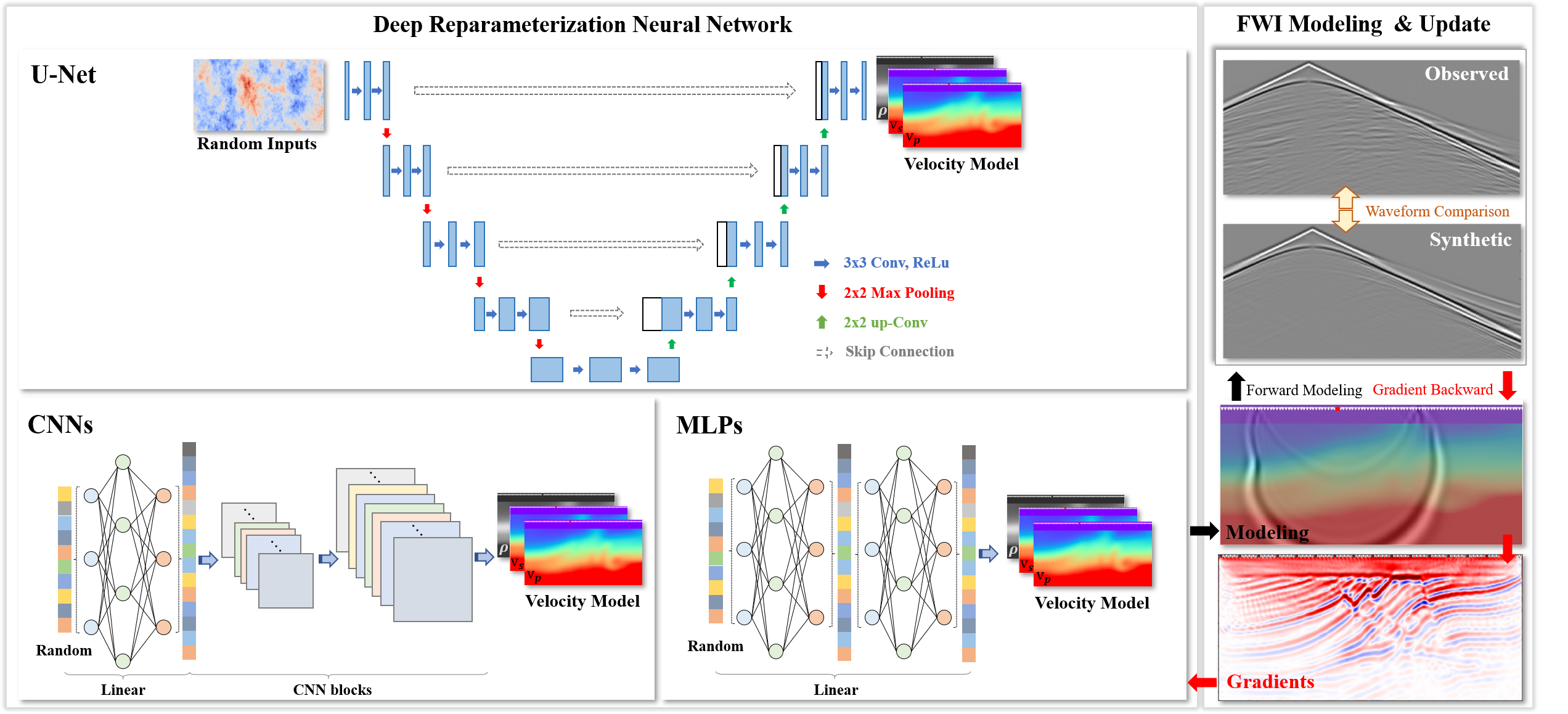}
    \caption{Workflow of deep reparameterization-based full waveform inversion. The left panel illustrates three types of reparameterization neural networks (U-Net, CNNs, and MLPs), which take random vectors or matrices as input and generate the corresponding subsurface models (e.g., $v_p$, $v_s$, and $\rho$). The right panel shows the FWI workflow, including forward modeling based on the generated velocity model, misfit computation between observed and synthetic waveforms, and gradient calculation using either the adjoint-state method or automatic differentiation.}
    \label{fig:deep_reparameterization_workflow}
\end{figure*}
    
\subsection{Deep Reparameterization-based FWI}

    Originally developed for image reconstruction, the deep image prior (DIP) was first introduced by Ulyanov et al.(2018) \cite{lempitsky_2018_Deep}. Their work demonstrated that the architecture of a generator network alone can capture a substantial amount of low-level image statistics, even prior to any learning process. Inspired by this insight, DIP has been successfully adapted to FWI by reparameterizing subsurface models via a generator network to capture hidden geostatistical features \cite{he_2021_Reparameterized, zhu_2021_Integrating, sun_2023_FullWaveform, wang_2023_Prior, liu_2025_Automatic}. In this framework (Fig.~\ref{fig:deep_reparameterization_workflow}), the deep reparameterization is formulated as
    \begin{equation}\label{eqn:deep_reparameterization}
        \mathbf{m} = \mathcal{N}(\theta | \textbf{I}_r),
    \end{equation}
    where $\theta$ denotes the parameters of the generator neural network $\mathcal{N}$, typically implemented using fully-connected neural networks (FCNNs), multi-layer convolutional neural networks (CNNs), or U-Net architectures. In the vanilla DIP formulation, the neural network input (often drawn from a normal distribution, i.e., \(\textbf{I}_r = \xi\) with \(\xi \sim \textit{Normal}(\mu,\sigma^2)\)) is randomly sampled and remains fixed throughout the optimization process \cite{lempitsky_2018_Deep}. With the reparameterized subsurface model, the optimization problem in Equation~\ref{eqn:objective_function} can be reformulated as 
    \begin{equation}\label{eqn:deep_reparameterization_objectivefunction}
        \textbf{m}^* = \underset{\textbf{m}}{\arg\min} \{ \mathcal{D}(d_{obs}, d_{cal}(\mathcal{N}(\theta | \textbf{I}_r))) + \lambda \mathcal{R}(\mathcal{N}(\theta | \textbf{I}_r))\}.
    \end{equation}

    Although vanilla DIP has shown great potential in image-related tasks such as reconstruction, denoising, inpainting, and super-resolution \cite{lempitsky_2018_Deep}, its direct application to FWI remains challenging due to the significantly higher complexity and nonlinearity of seismic inversion. Consequently, relatively few studies have directly applied the vanilla DIP formulation to FWI. A conceptually related strategy is physics-informed implicit FWI (IFWI), which employs a FCNN to map spatial coordinates ($x_i, z_i$) to the corresponding local velocity model values (\(\textbf{m}_{x_i, y_i, z_i}\)). The network output is then integrated into the conventional FWI workflow \cite{sun_2023_Implicit}, and the gradient of the conventional FWI objective function with respect to the model parameters, (\(\partial \mathcal{L}/ \partial \textbf{m}\)), is linked to the gradient of the network output with respect to its parameters, (\(\partial \textbf{m} / \partial \theta\)), through the chain rule of automatic differentiation:
    \begin{equation}\label{eqn:DIP_gradient}
        \frac{\partial \mathcal{L}}{\partial \theta} = \frac{\partial \mathcal{L}}{\partial \textbf{m}} . \frac{\partial \textbf{m}}{\partial \theta} = \frac{\partial \mathcal{L}}{\partial \mathcal{N}(\theta | \textbf{I}_r)} . \frac{\partial \mathcal{N}(\theta | \textbf{I}_r)}{\partial \theta}.
    \end{equation}
    where, \( \partial \mathcal{N}(\theta | \textbf{I}_r) / \partial \theta\) represents the Jacobian matrix of the network with respect to $\theta$, which can be computed using automatic differentiation. The second gradient item, \(\partial \mathcal{L} / \partial \mathcal{N}(\theta | \textbf{I}_r)\), is the gradient of FWI objective with respect to $\textbf{m}$, which can be obtained through adjoint-state methods or automatic differentiation.
    
    Synthetic experiments have shown that IFWI can produce acceptable inversion results even when initialized with random inputs, thereby significantly reducing the traditional FWI’s dependence on accurate initial velocity models. However, the authors acknowledged that IFWI typically requires thousands of iterations to achieve satisfactory results, and its performance is highly sensitive to the network architecture and parameter tuning, especially under varying geological conditions. 
    
    In practical applications, a preliminary velocity model is often available (e.g., a smoothed model from travel-time tomography). Leveraging such prior information can substantially reduce the number of required inversion iterations and mitigate the challenges associated with hyperparameter selection. This insight motivates the development of reference-guided DIP, which incorporates prior knowledge from the initial velocity model to improve inversion efficiency and stability \cite{zhao_2020_Referencedriven}.

\subsection{Reference-Guided Deep Reparameterization}

    Leveraging prior information, the reference-guided DIP approach replaces the random input of the vanilla DIP with a given reference \cite{zhao_2020_Referencedriven}. Although the original reference-guided DIP proposed by Zhao et al. (2020) \cite{zhao_2020_Referencedriven} exhibits improved performance over the vanilla DIP by incorporating an initial solution closer to the final result, the network is still randomly initialized and must be learned from scratch. Consequently, this approach typically requires a large number of iterations (often thousands), and given that each iteration involves a full FWI calculation, the computational cost is substantial. 
    
    To address these issues and further improve inversion results, we propose and compare several strategies for integrating the reference (initial) velocity model into the deep reparameterization framework:
    \begin{enumerate}
        \item \textbf{Direct Input of the Reference Model \cite{zhao_2020_Referencedriven}:} The reference subsurface model \(m_0\) is used as the input to the generator neural network, and the reparameterization subsurface model is expressed as:
        \begin{equation}\label{eqn:strategy1}
            \textbf{m} = \mathcal{N}(\theta \mid \textbf{m}_0).
        \end{equation}
        \item \textbf{Perturbation Learning \cite{zhu_2021_Integrating}:} The generator neural network takes random noise $\xi$ as input and learns to predict the perturbation relative to the reference subsurface model. The velocity model used in the FWI workflow is obtained by summing the network output with the reference velocity model
        \begin{equation}\label{eqn:strategy2}
            \textbf{m} = \mathcal{N}(\theta \mid \xi) + \textbf{m}_0 .
        \end{equation}
        \item \textbf{Pre-training (Warm-Up) with Reference model \cite{he_2021_Reparameterized}:} The generator neural network is first pre-trained to learn a mapping from random noise to the reference velocity model, expressed as 
        \begin{equation}\label{eqn:strategy3}
            \theta^* = \underset{\theta}{\arg\min}{(\| \mathcal{N}(\theta \mid \xi) - \textbf{m}_0 \|_2^2)}.
        \end{equation}
        Following this warm-up phase, the pretrained network is used in the FWI process to generate the input velocity model, i.e., $\textbf{m} = \mathcal{N}(\theta^* \mid \xi)$, and further updated during inversion.
    \end{enumerate}

    Most existing DIP-FWI studies focus on the regularization effects under noisy conditions \cite{zhu_2021_Integrating, sun_2023_Implicit, liu_2025_Automatic}, while its broader potential for addressing practical FWI challenges remains underexplored. Notably, the problem of missing data in seismic acquisition shares structural similarity with image inpainting \cite{lempitsky_2018_Deep}. In this work, we further investigate the effectiveness of reference-guided DR-FWI in addressing missing data scenarios, including sparse shot-gathers and missing traces. 

    In the following sections, we conduct a detailed comparative analysis of these proposed strategies, focusing on their inversion performance and convergence behavior. Furthermore, we assess their robustness in handling missing data scenarios, demonstrating the advantages of proposed approach in improving inversion results under challenging conditions.

\subsection{Deep Reparameterization for multiparameter FWI}

    The success of deep reparameterization in single-parameter inversion has motivated its extension to multiparameter FWI. One of the primary challenges for multiparameter FWI is crosstalk, which arises when data residuals caused by inaccuracies in one physical parameter are misattributed to another \cite{operto_2013_Guided}. This phenomenon hinders convergence and degrades inversion accuracy. Traditional solutions to mitigating crosstalk typically involve weighting updates for different physical parameters or post-processing gradients using Hessian-based corrections, which are either complex or hard to implement.

    In contrast, deep reparameterization provides an alternative framework in which multiple physical parameters (e.g., $v_p$ and $\rho$ in acoustic wave equations, $v_p$, $v_s$, and $\rho$ in elastic wave equations) are simultaneously represented by a generative neural network. Under the sparsity constraints of the network’s parameter space, these physical parameters can be automatically coupled, reducing the need for explicit crosstalk suppression:
    \begin{equation}\label{eqn:multiparam_reparameterization}
        \mathbf{m_1},\dots,\mathbf{m_n} = \mathcal{N}(\theta | \textbf{I}_r).
    \end{equation}
    
\begin{figure}[!ht]
        \centering
        \includegraphics[width=0.5\textwidth]{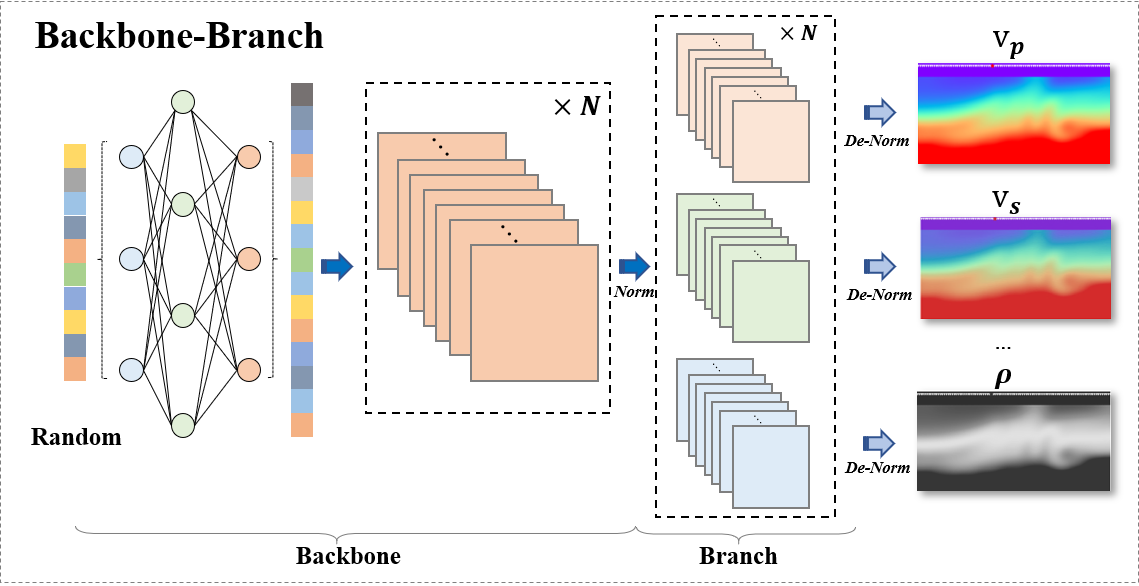}
        \caption{Overview of the "backbone-branch" architecture for multiparameter reparameterization. Random inputs are first projected into a high-dimensional latent space via a linear layer to form an expressive representation. The shared backbone, composed of multiple convolutional blocks, is designed to extract common structural features across different physical parameters. Each branch, also composed of convolutional blocks, is designed to capture parameter-specific variations. Normalization and denormalization operations adapt to different physical scales.}
        \label{fig:multi_parameter_architecture}
    \end{figure}

    Experimental results indicate that the coupling problem in multiparameter inversion can be significantly mitigated using a simple multi-layer CNN with a "backbone-branch" architecture (Fig.~\ref{fig:multi_parameter_architecture}), given an appropriate parameterization strategy. Two key strategies include 1) \textbf{Hierarchical Representation}: A backbone network captures shared structural features and facilitate automatic coupling among different physical parameters, while separate branch networks learn parameter-specific variations. 2) \textbf{Parameter Normalization}: Normalizing the output of each branch network stabilize the inversion process, while subsequent inverse normalization ensures that the predictions remain consistent with the physical scales of each parameter. By integrating these strategies, the proposed deep reparameterization framework improves the stability of multiparameter inversion while reducing the reliance on traditional crosstalk correction techniques.

%
%

\section{Experiments and Results}

\begin{table*}[!ht]
\centering
\renewcommand \arraystretch{1.5}
\caption{Key Parameters of the Tests for DR-FWI in Various Types of Earth Models}
\begin{tabular}{lccccc}
\hline
                                           & \multicolumn{1}{l}{\textbf{Acoustic Marmousi2}} & \multicolumn{1}{l}{\textbf{Acoustic Overthrust}} & \multicolumn{1}{l}{\textbf{Acoustic FootHill}} & \multicolumn{1}{l}{\textbf{Elastic Anomaly}} & \multicolumn{1}{l}{\textbf{Elastic Marmousi2}} \\ \hline
\multicolumn{1}{l|}{\textbf{Number of grids}}       & 76$\times$200                          & 90$\times$200                           & 97$\times$267                         & 100$\times$200                      & 68$\times$200                         \\
\multicolumn{1}{l|}{\textbf{Grid Spacing}}          & 40 m                                   & 50 m                                    & 40 m                                  & 30 m                                & 45 m                                  \\
\multicolumn{1}{l|}{\textbf{Time Steps}}            & 2500                                   & 1600                                    & 2000                                  & 2500                                & 2500                                  \\
\multicolumn{1}{l|}{\textbf{Time Interval}}         & 0.003 s                                & 0.003 s                                 & 0.003 s                               & 0.0025 s                            & 0.003 s                               \\
\multicolumn{1}{l|}{\textbf{Number of Receivers}}   & 200                                    & 200                                     & 267                                   & 200                                 & 200                                   \\
\multicolumn{1}{l|}{\textbf{Number of Sources}}     & 40                                     & 40                                      & 54                                    & 40                                  & 40                                    \\
\multicolumn{1}{l|}{\textbf{Source dominant freq.}} & 5 Hz                                   & 5 Hz                                    & 5 Hz                                  & 5 Hz                                & 5 Hz                                  \\
\multicolumn{1}{l|}{\textbf{Smooth window size}}    & 240 m$\times$240 m                     & 300 m$\times$300 m                      & 240 m$\times$240 m                    & -                                   & 180 m$\times$180 m                    \\
\multicolumn{1}{l|}{\textbf{Inverted parameters}}   & $v_p$                                  & $v_p$                                   & $v_p$                                 & $v_p$,$v_s$,$\rho$                  & $v_p$,$v_s$,$\rho$                    \\ \hline
\end{tabular}
\label{tab:velmodel_introduction}
\end{table*}

\subsection{Baseline Models and Quantitative Metrics}
    To evaluate the effectiveness of reference-guided deep reparameterization, we select three well-known 2D geophysical models: the Marmousi2 model \cite{martin_2006_Marmousi2}, the Overthrust model \cite{aminsadeh_1996_3D}, and the Foothill model \cite{gray_1995_Migration}, the corresponding initial velocity model is obtained by applying a Gaussian smoothing operation. More details about these models and the corresponding observed system definitions can be found in Table~\ref{tab:velmodel_introduction}. All experiments are conducted on a single NVIDIA GPU and implemented using PyTorch. Forward modeling for acoustic and elastic wave propagation is performed using the ADFWI toolbox \cite{liu_2024_Liufeng2317}. The global-correlation misfit function is adopted to measure the discrepancy between the observed and synthetic waveforms \cite{choi_2012_Application}, and the Adam optimizer is employed to iteratively update the velocity model and network parameters \cite{kingma_2017_Adam}. For optimization, we tune the learning rate and adopt the StepLR learning rate decay strategy, which decreases the initial learning rate by a decay factor every $N$ iterations. The maximum number of iterations for inversion is set to 300.

    To assess the inversion performance of different methods, we employ the mean absolute percentage error (MAPE) \cite{hyndman_2006_Another}, the structural similarity index measure (SSIM) \cite{wang_2004_Image}, and the signal-to-noise ratio (SNR) to quantify the discrepancy between the inverted and true models:
    \begin{equation}\label{eqn:metric_mape}
        MAPE(v, \hat{v}) = \frac{1}{m \times n} \sum_{i = 1}^{m} \sum_{j = 1}^{n} \left| \frac{v_{i,j} - \hat{v}_{i,j}}{v_{i,j}} \right| \times 100\%,
    \end{equation}
    \begin{equation}\label{eqn:metric_ssim}
        SSIM(v, \hat{v}) = \frac{(2 \mu_{v} \mu_{\hat{v}} + c_1)(2 \sigma_{v\hat{v}} + c_2)}{(\mu_{v}^2 + \mu_{\hat{v}}^2 + c_1)(\sigma_{v}^2 + \sigma_{\hat{v}}^2 + c_2)},
    \end{equation}
    \begin{equation}\label{eqn:metric_snr}
        SNR(v, \hat{v}) = 10 log_{10} \frac{\| v \|_2^2}{\| \hat{v} - v \|_2^2},
    \end{equation}
    where \(m\) and \(n\) are velocity model size, \(v_{i,j}\) and \(\hat{v}_{i,j}\) are the true and inverted velocity models, respectively, \(\mu_{v}\) and \(\sigma_{v}\) are the local mean and standard deviation for the true model, \(\mu_{\hat{v}}\) and \(\sigma_{\hat{v}}\) are those for the inverted model, and \(\sigma_{v,\hat{v}}\) is the cross-covariance between the true and inverted models. Higher SSIM and SNR values, along with lower MAPE, indicate better inversion performance.

\subsection{Ablation Studies on Reparameterization Network Architectures and Strategies}
    Previous studies have demonstrated the potential of deep reparameterization networks in FWI, with varying architectures and strategies exhibiting distinct levels of performance. Commonly adopted architectures include MLPs, CNNs, and U-Net, while two prevalent reparameterization strategies involve learning the velocity perturbation ($\Delta\mathbf{m}$, Equation~\ref{eqn:strategy2}) or directly initializing the neural network to learn the velocity model ($\mathbf{m}$, Equation~\ref{eqn:strategy3}). 
    
    A fundamental question thus arises: \textbf{which combination of reparameterization network architecture and training strategy yields the most effective FWI performance?} \cite{he_2021_Reparameterized}. To address this, we evaluate three distinct neural network architectures: 
    \begin{enumerate}
        \item \textbf{U-Net}: The input is a randomly initialized 2-D arrays, from which the network learns the statistical representation of the velocity model through a sequence of down-sampling and up-sampling operations. Skip connections are incorporated to retain fine-grained details by linking corresponding layers in the down-sampling and up-sampling paths. 
        \item \textbf{MLPs}: The input is a 1-D random vector, which is transformed into the velocity model through multiple fully connected layers.
        \item \textbf{CNNs}: The input consists of 1-D random arrays, which are first processed through a fully connected layer to generate an initial 1-D representation of the velocity model. The representation is then reshaped into a 2-D format and refined using multiple convolutional layers. Finally, a $1\times1$ convolutional layer merges multi-channel representations into the final velocity model output.
    \end{enumerate}

\begin{figure}[!ht]
        \centering
        \includegraphics[width=0.5\textwidth]{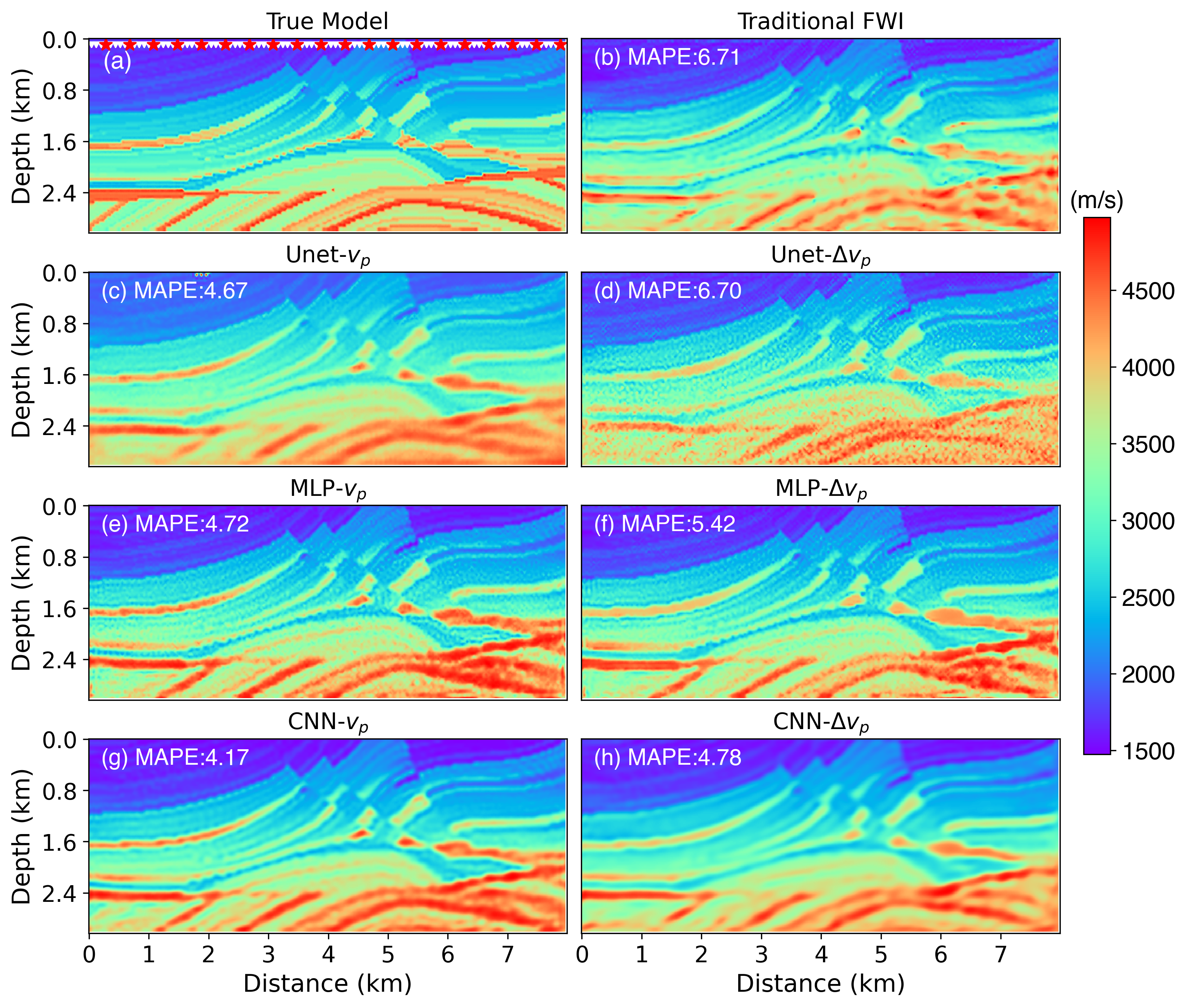}
        \caption{Comparison of inversion results using different deep reparameterization architectures and strategies for the Marmousi2 model. (a) True velocity model. (b) Baseline inversion result using conventional FWI. (c)-(d) Results obtained using U-Net-based reparameterization for $v_p$ and $\Delta v_p$. (e)-(f) Results obtained using MLP-based reparameterization for $v_p$ and $\Delta v_p$. (g)-(h) Results obtained using CNN-based reparameterization for $v_p$ and $\Delta v_p$.}
        \label{fig:architecture_and_strategy_marmousi2_model}
    \end{figure}

    Combining the above architectures with the two deep reparameterization strategies leads to six different combinations: CNN-$v_p$, CNN-$\Delta v_p$, MLP-$v_p$, MLP-$\Delta v_p$, U-Net-$v_p$, and U-Net-$\Delta v_p$. We first evaluate all six combinations on the Marmousi2 model. To reduce the impact of network hyperparameter choices (e.g., the number of layers and channels), we perform a network architecture search for each architecture and select the best-performing configuration for comparison. Specifically, we explore CNNs architectures with 1 to 4 layers, each with 64, 128, 256, and 512 channels, totaling 16 different variants. For MLPs, we test 1 and 2 layers architectures, each with 100, 1000, 2000, and 3000 neurons per layer, resulting in 8 variants. For U-Net, we test 1 to 4 encoder–decoder blocks, each with 64, 128, 256, and 512 channels, yielding 16 variants. The best inversion results for each reparameterization strategy on the Marmousi2 model are illustrated in Fig.~\ref{fig:architecture_and_strategy_marmousi2_model}. Panel (a) depicts the true Marmousi2 model ($v_p$), while panel (b) presents the inversion result using traditional FWI. Panels (c) and (d) correspond to the U-Net-based inversion results using the $v_p$ and $\Delta v_p$ initial model embedding strategies, respectively. Panels (e) and (f) show the results for the MLP-based variants, respectively. Panels (g) and (h) depict the CNN-based results.
    
\begin{figure}[!ht]
        \centering
        \includegraphics[width=0.48\textwidth]{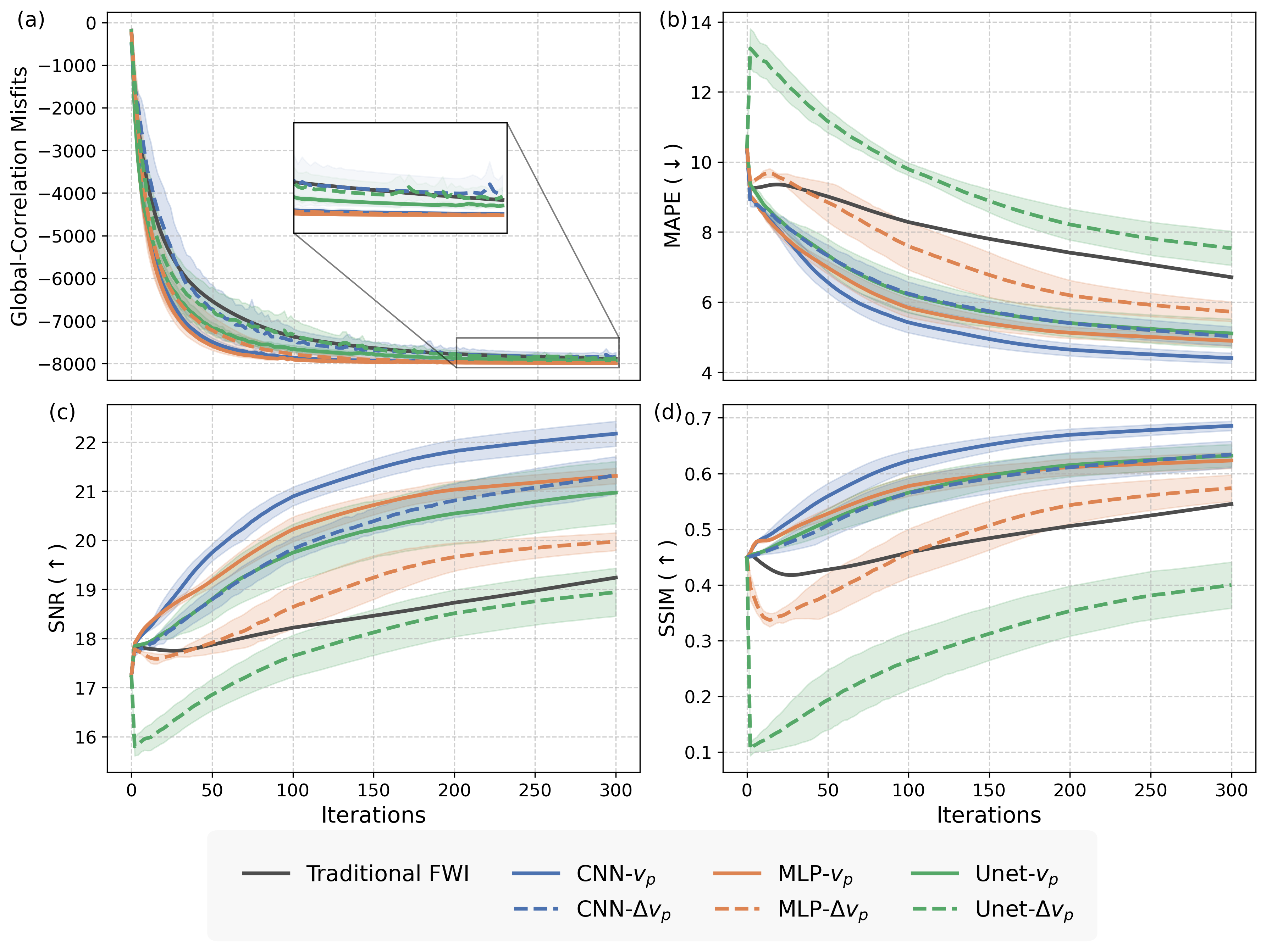}
        \caption{Comparison of inversion performance metrics for different deep reparameterization architectures and strategies. (a) Evolution of the global-correlation misfit. (b)-(d) Model residuals in terms of MAPE, SNR, and SSIM, respectively. Solid (CNN-$v_p$, MLP-$v_p$, Unet-$v_p$) and dashed (CNN-$\Delta v_p$, MLP-$\Delta v_p$, Unet-$\Delta v_p$) lines represent the mean performance for each reparameterization strategy, while the shaded areas indicate variations induced by different hyperparameter choices (e.g., the number of layers and channels).}
        \label{fig:architecture_and_strategy_marmousi2_misfit}
    \end{figure}

    To further assess the inversion performance of different strategies, we analyze the evolution of data misfit and model residuals, as shown in Fig.~\ref{fig:architecture_and_strategy_marmousi2_misfit}. The colored lines represent the average performance across different network variants, while the shaded regions indicate the range of fluctuations in inversion results due to hyperparameter differences. Panel (a) presents the evolution of the global-correlation misfit (GC-loss), demonstrating that all inversion strategies successfully converge in the data domain. Panels (b), (c), and (d) show model residuals in terms of MAPE, SNR, and SSIM, respectively. The best performance achieved by each reparameterization strategy is summarized in Table~\ref{tab:marmousi2_overthrust_foothill_metric}. Two key findings emerge from these results: 1) The warm-up strategy (CNN-$v_p$, MLP-$v_p$, U-Net-$v_p$) consistently outperforms the perturbation-learning strategy (CNN-$\Delta v_p$, MLP-$\Delta v_p$, U-Net-$\Delta v_p$) across all network architectures. 2) A simple CNN-based reparameterization network achieves superior inversion performance compared to the baseline and even outperforms more complex architectures such as U-Net. Beyond the Marmousi2 model, we conducted similar experiments on the Overthrust and Foothill models. The quantitative results for these datasets are summarized in Table~\ref{tab:marmousi2_overthrust_foothill_metric}.

\begin{table*}[!ht]
\caption{Quantitative comparison of reparameterization strategies on the Marmousi2 model based on GC-loss, MAPE, SSIM, and SNR.}
\centering
\renewcommand \arraystretch{1.5}
\begin{tabular}{lccccccc}
\hline
                                          & \textbf{Traditional FWI} & \textbf{Unet-$\Delta v_p$} & \textbf{MLP-$\Delta v_p$} & \textbf{CNN-$\Delta v_p$} & \textbf{Unet-$v_p$} & \textbf{MLP-$v_p$} & \textbf{CNN-$v_p$}      \\ \hline
\multicolumn{1}{l|}{\textbf{GC-loss (Marmousi2)}}  & -7892.49        & -7937.52          & -7989.94         & -7987.98         & -7944.44   & -7991.62  & -7990.07       \\
\multicolumn{1}{l|}{\textbf{MAPE ($\downarrow$)}}  & 6.71            & 6.70              & 5.42             & 4.78             & 4.67       & 4.72      & \textbf{4.17}  \\
\multicolumn{1}{l|}{\textbf{SSIM ($\uparrow$)}}    & 0.55            & 0.44              & 0.60             & 0.66             & 0.66       & 0.64      & \textbf{0.70}  \\
\multicolumn{1}{l|}{\textbf{SNR ($\uparrow$)}}     & 19.24           & 19.38             & 20.24            & 21.95            & 21.73      & 21.54     & \textbf{22.57} \\ \hline
\multicolumn{1}{l|}{\textbf{GC-loss (Overthrust)}} & -7957.52        & -7985.95          & -7993.14         & -7984.25         & -7984.37   & -7993.53  & -7991.66       \\
\multicolumn{1}{l|}{\textbf{MAPE ($\downarrow$)}}  & 3.91            & 4.43              & 2.99             & 3.05             & 2.75       & 2.52      & \textbf{2.38}  \\
\multicolumn{1}{l|}{\textbf{SSIM ($\uparrow$)}}    & 0.66            & 0.45              & 0.63             & 0.74             & 0.75       & 0.76      & \textbf{0.81}  \\
\multicolumn{1}{l|}{\textbf{SNR ($\uparrow$)}}     & 25.83           & 25.08             & 28.21            & 27.83            & 28.46      & 29.41     & \textbf{29.52} \\ \hline
\multicolumn{1}{l|}{\textbf{GC-loss (Foothill)}}   & -14388.72       & -14378.79         & -14379.05        & -14366.12        & -14407.45  & -14410.76 & -14407.59      \\
\multicolumn{1}{l|}{\textbf{MAPE ($\downarrow$)}}  & 1.63            & 3.34              & 2.73             & 1.82             & 1.41       & 1.41      & \textbf{1.37}  \\
\multicolumn{1}{l|}{\textbf{SSIM ($\uparrow$)}}    & 0.84            & 0.50              & 0.57             & 0.84             & 0.85       & 0.82      & \textbf{0.87}  \\
\multicolumn{1}{l|}{\textbf{SNR ($\uparrow$)}}     & 25.03           & 26.17             & 27.68            & 30.34            & 32.20      & 32.25     & \textbf{32.32} \\ \hline
\end{tabular}
\label{tab:marmousi2_overthrust_foothill_metric}
\end{table*}
    
\subsection{Robustness Analysis under Noise and Acquisition Sparsity}
    
    Neural networks are known to exhibit a spectral bias during the learning process in the frequency domain, prioritizing low-frequency signal components before gradually resolving high-frequency details \cite{rahaman_2018_Spectral, chakrabarty_2019_Spectral}. This property is aligned with the multi-scale principles of FWI, where low-frequency data constrain large-scale structures prior to high-frequency refinements. In deep reparameterized FWI, the neural network progressively learns statistical characteristics of the velocity model through iterative inversion, enhancing its structural representation capability. The spectral bias inherently imposes a prior constraint, reducing the risk of local minima entrapment by initially recovering smooth velocity models (low-frequency dominance) and later refining high-frequency details. This regularization mechanism not only stabilizes the inversion workflow but also improves robustness against noise and data sparsity. To validate these claims, rigorous evaluations are conducted using the Marmousi2 model under controlled noise and sparsity conditions.

\begin{figure}[!ht]
        \centering
        \includegraphics[width=0.5\textwidth]{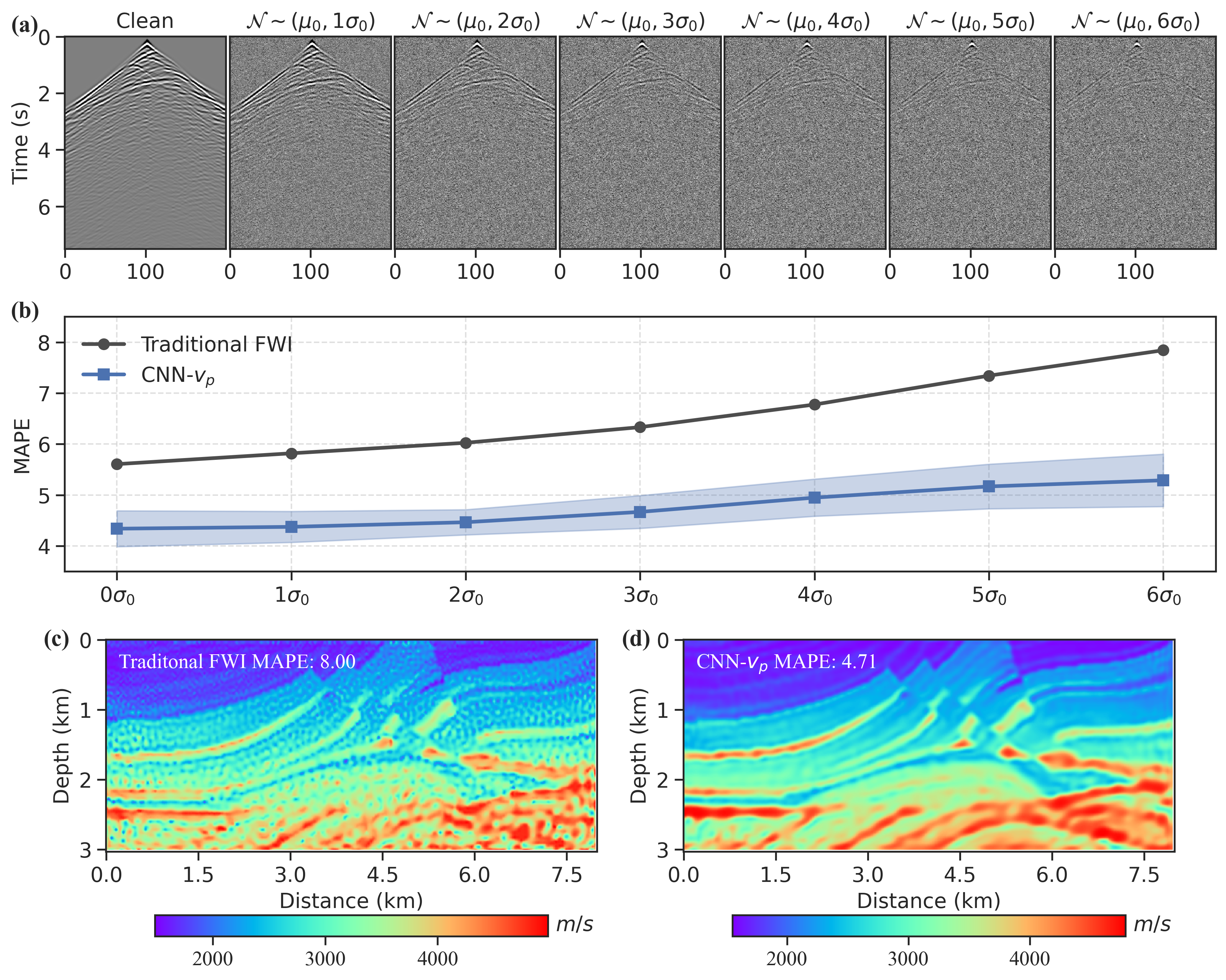}
        \caption{Robustness evaluation of deep reparameterized FWI under noisy data scenarios on the Marmousi2 model. (a) Example shot gather showing clean and noisy data with Gaussian noise levels from $ 1\sigma_0$ to $6\sigma_0$. (b) MAPE of inversion results with and without the CNN-$v_p$ deep reparameterization strategy under different noise conditions. Shaded areas represent MAPE variations across different CNN architectures. (c) and (d) Inversion results with $6\sigma_0$ Gaussian noise, highlighting the improved stability of the deep reparameterization method under high noise levels.}
        \label{fig:noise_robust_test}
    \end{figure}

    First, noise robustness of the deep reparameterization method is quantified by injecting Gaussian white noise into observation data. The noise levels are defined by a fixed mean $\mu = \mu_0$ and varying standard deviations from $\sigma = 1\sigma_0$ to $\sigma = 6\sigma_0$, where $\mu_0$ and $\sigma_0$ correspond to the mean and standard deviation of the clean observation data, respectively. Fig.~\ref{fig:noise_robust_test}a compares a clean shot gather with its noise-contaminated counterpart. Building on prior results demonstrating the superiority of the CNN-$v_p$ architecture (CNN-based velocity parameterization with pretraining-based initialization), we adopt this strategy for the noise robustness study. Fig.~\ref{fig:noise_robust_test}b presents the MAPE of inversion results across increasing noise levels: conventional FWI (black line) versus CNN-$v_p$ reparameterization (blue line), with shaded regions indicating architectural variability across CNNs. Figs.~\ref{fig:noise_robust_test}c and d visualize inversion results for the $6\sigma_0$ noise case, highlighting the enhanced stability and accuracy of the reparameterized approach under extreme-noise conditions.

\begin{figure}[!ht]
        \centering
        \includegraphics[width=0.5\textwidth]{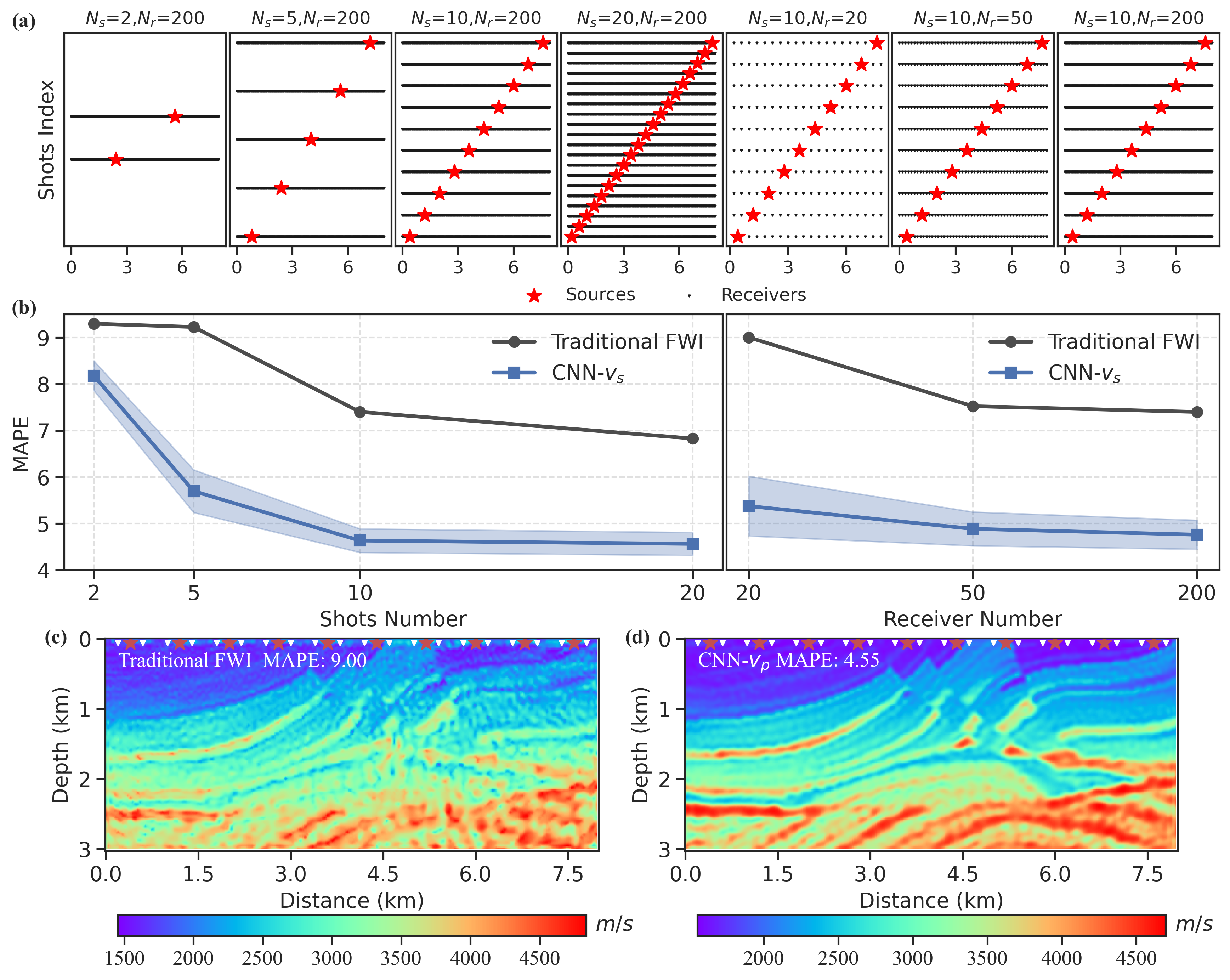}
        \caption{Robustness evaluation of deep reparameterized FWI under sparse-data scenarios on the Marmousi2 model. (a) Seven observation configurations: the first four fix 200 receivers with source counts ranging from 2 to 20; the last three fix 10 sources with receiver counts increasing from 20 to 200. (b) MAPE curves comparing inversion accuracy for sparse-source (left: 2 to 20 sources, 200 receivers) and sparse-receiver (right: 10 sources, 20 to 200 receivers) scenarios, with (blue line) and without (black line) deep reparameterization. Shaded regions indicate MAPE variability across CNN architectures. (c) Conventional FWI results and (d) deep reparameterized FWI results for an extreme receiver-sparse configuration (10 sources, 20 receivers). }
        \label{fig:imputation_robust_test}
    \end{figure}

    While the robustness of deep reparameterized FWI to noise has been extensively validated \cite{zhu_2021_Integrating, sun_2023_Implicit}, its effectiveness under data sparsity remains less well explored. To evaluate this aspect, we design two sparsity scenarios: (1) sparse sources (20, 10, 5, and 2 sources with 200 receivers); and (2) sparse receivers (10 sources with 200, 50, or 20 receivers). Fig.~\ref{fig:imputation_robust_test}a summarizes seven acquisition configurations: four fixed-receiver systems (200 receivers, 2 to 20 sources) and three fixed-source systems (10 sources, 20 to 200 receivers). Fig.~\ref{fig:imputation_robust_test}b presents MAPE curves for both the sparse-source (left) and the sparse-receiver (right) cases, comparing conventional and reparameterized FWI. Figs.~\ref{fig:imputation_robust_test}c and d compare the inverted velocity models for an extreme receiver-sparse case (10 sources, 20 receivers), highlighting the reparameterized method’s superior structural recovery (Fig.~\ref{fig:imputation_robust_test}d). Overall, under non-extreme source sparsity conditions ($>$ 2 sources), the reparameterized FWI exhibits significantly higher stability and accuracy compared to conventional FWI. Even under extreme sparsity regimes (2 sources or 20 receivers), it retains a measurable performance advantage over conventional approaches, with particularly pronounced benefits observed in receiver-sparse configurations (Figs.~\ref{fig:imputation_robust_test}d).
    
\subsection{Multiparameter Inversion and Crosstalk Mitigation}
        
    The superior performance of deep reparameterized FWI in single-parameter inversion stems from the inherent spectral bias and implicit prior regularization of deep neural networks during training. Beyond these properties, the flexibility of this framework enables seamless extension to multiparameter inversion by incorporating a backbone–branch structure. In multiparameter reparameterized FWI, a shared backbone network captures coupled representations of subsurface parameters (e.g., $v_p$, $v_s$, $\rho$), while dedicated branch networks disentangle parameter-specific features through channel-wise normalization tailored to their respective physical ranges. This design is validated through two synthetic experiments: a geometrically idealized anomaly model and the Marmousi2 benchmark.

\begin{figure}[!ht]
        \centering
        \includegraphics[width=0.5\textwidth]{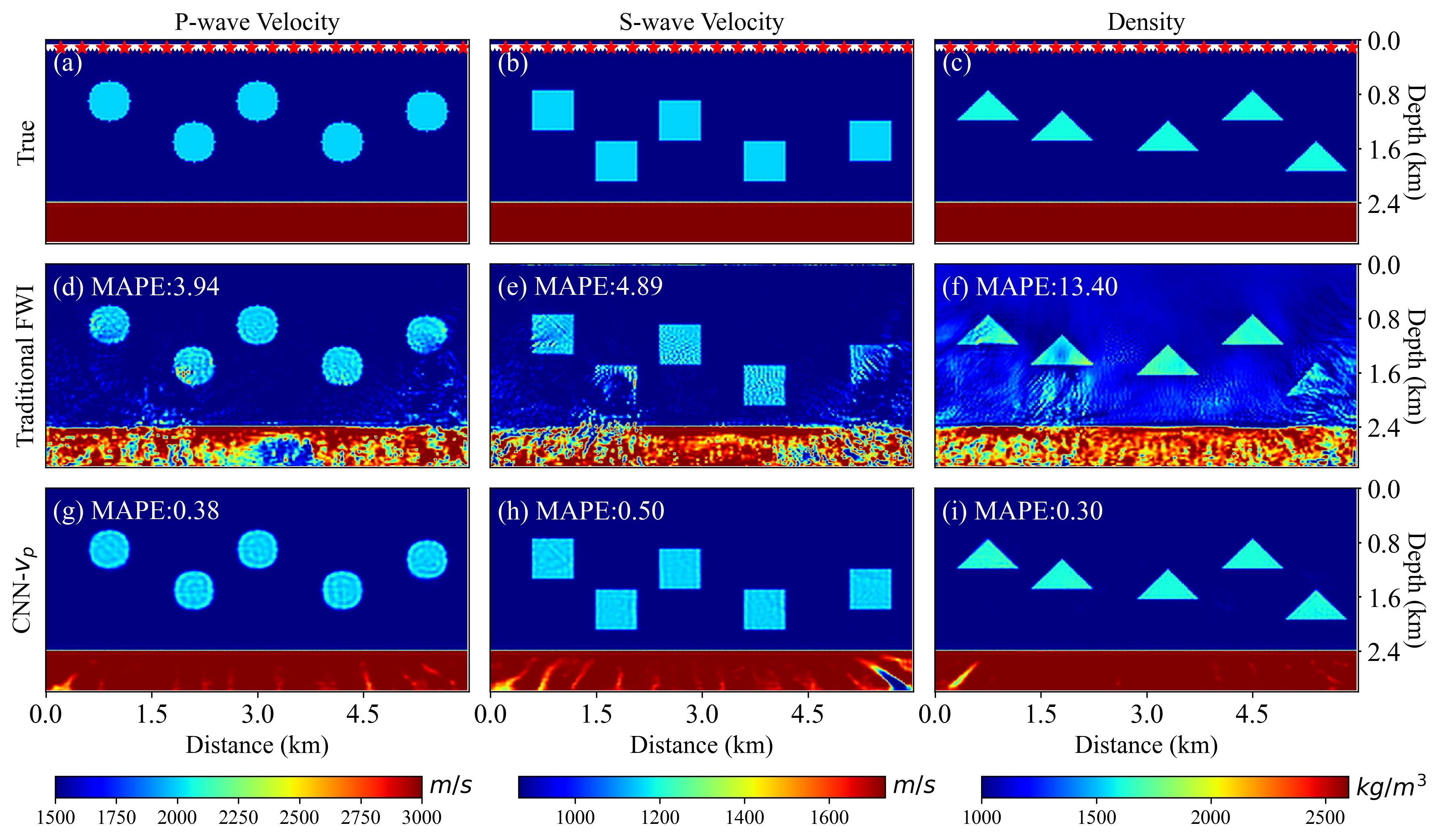}
        \caption{Multiparameter inversion results on an anomaly model. (a–c) Ground-truth models of $v_p$, $v_s$, and $\rho$, respectively. (d–f) Inversion results obtained using traditional FWI. (g–i) Inversion results obtained using DR-FWI. Red stars and white triangles denote source and receiver locations, respectively.}
        \label{fig:multiparameter_anomaly_test}
    \end{figure}

    To evaluate the crosstalk mitigation capabilities of the proposed multiparameter inversion framework, we first design a synthetic elastic model inspired by the CTS model \cite{dokter_2017_Full} and STH model \cite{dhara_2022_ElasticAdjointNet}. The designed model consists of an elastic layer overlying a homogeneous half-space and embeds three distinct types of geometric anomalies: five circular $v_p$ anomalies, five square $v_s$ anomalies, and five triangular $\rho$ anomalies. Despite their idealized nature, these non-overlapping and geometrically distinct anomalies (Fig.~\ref{fig:multiparameter_anomaly_test}a-c) provide a controlled testbed for isolating and quantifying crosstalk effects. The model spans a $100 \times 200$ grid with 30-meter spacing, using 40 sources and 200 receivers. We perform simultaneous inversion for $v_p$, $v_s$ and $\rho$—a parameter triplet commonly adopted in elastic FWI. All inversions are initialized from homogeneous background models devoid of any embedded anomalies. The results of conventional FWI (Fig.~\ref{fig:multiparameter_anomaly_test}d-f) reveal severe crosstalk artifacts: $v_p$ and $\rho$ models show spurious structures mirroring the $\rho$ anomalies, while the $\rho$ model, despite preserving the intended geometry, suffers from unstable background regions (MAPE \(>\) 13). In contrast, the CNN-$v_p$ reparameterized FWI (Fig.~\ref{fig:multiparameter_anomaly_test}g-i) achieves near-perfect recovery of all three parameters (MAPE \(<\) 0.5), eliminating crosstalk and preserving background stability. These results demonstrate that the proposed deep reparameterization framework effectively decouples multiparameter interactions while enforcing physically consistent relationships across multiple elastic properties.

\begin{figure}[!ht]
        \centering
        \includegraphics[width=0.5\textwidth]{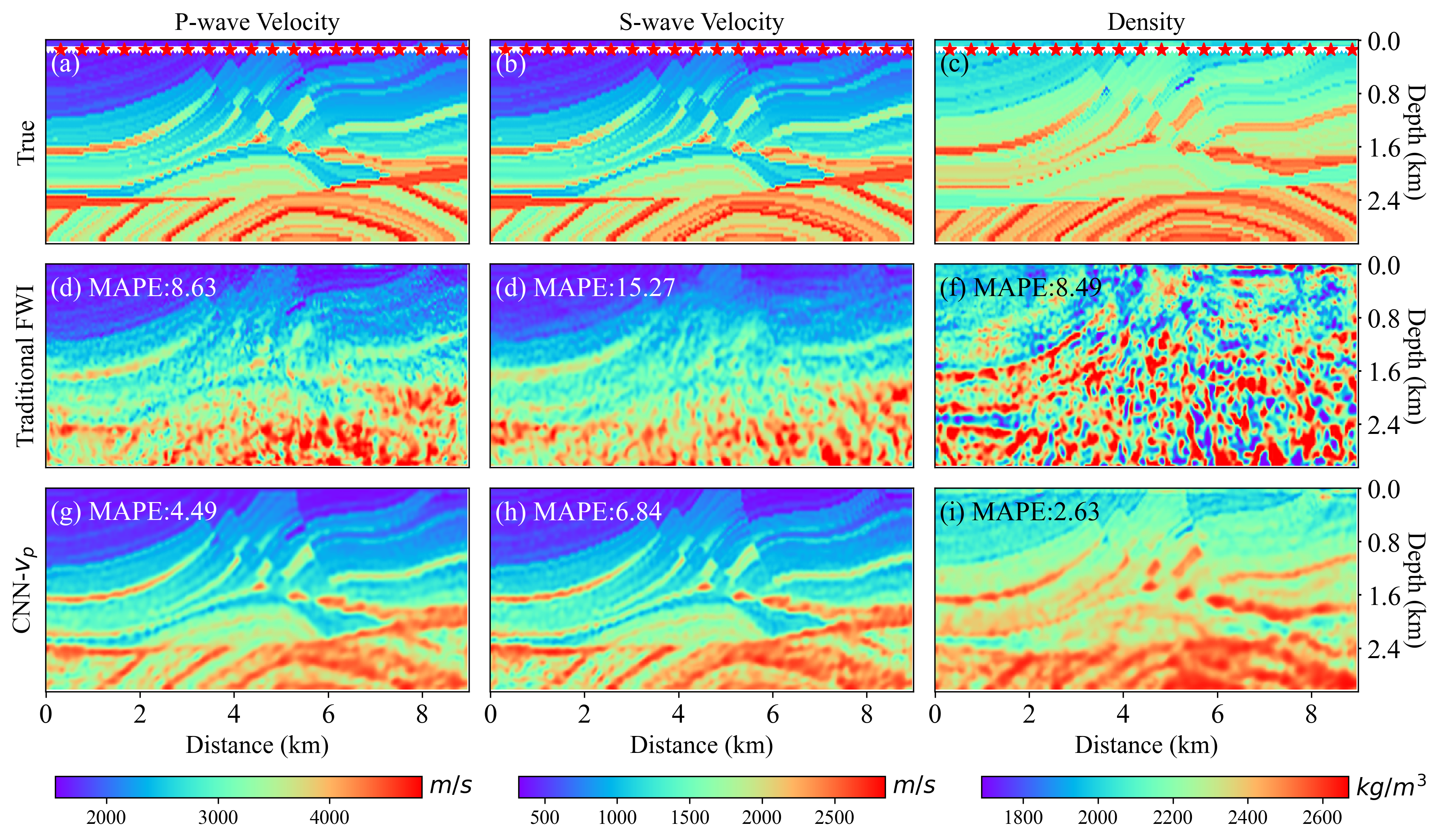}
        \caption{Multiparameter inversion on Marmousi2 model. (a–c) Ground-truth models of $v_p$, $v_s$, and $\rho$, respectively. (d–f) Inversion results obtained using traditional FWI. (g–i) Inversion results obtained using DR-FWI. Red stars and white triangles denote source and receiver locations, respectively.}
        \label{fig:multiparameter_marmousi2_test}
    \end{figure}

    Building upon the validated effectiveness of deep reparameterization in suppressing crosstalk in the simplified anomaly model, we further apply this approach to the more geologically realistic Marmousi2 model. The true $v_p$, $v_s$, and $ \rho$ models are shown in Figs.~\ref{fig:multiparameter_marmousi2_test}a–c. Figs.~\ref{fig:multiparameter_marmousi2_test}d–f display the results obtained by conventional FWI without reparameterization. Due to strong parameter coupling, the deeper regions of the $v_p$ and $v_s$ models exhibit significant blurring, while the $\rho$ model, being less sensitive to the seismic data, appears structurally incoherent and highly disordered. In contrast, the inversion results obtained by the proposed deep reparameterization method (Figs.~\ref{fig:multiparameter_marmousi2_test}g–i) show substantial improvements. All three parameters are recovered with enhanced structural fidelity and reduced artifacts, closely matching the ground truth (Fig.~\ref{fig:multiparameter_marmousi2_test}a–c). The combination of deep regularization and effective crosstalk suppression enables stable and accurate recovery of complex subsurface properties. This results highlight the robustness and generalizability of the proposed reparameterization framework in complex geological settings, further confirming its applicability to multiparameter inversion tasks.

%
%

\section{DISCUSSION}

\subsection{Insights on Deep Reparameterization and DIP Mechanism}

    Although empirical results across multiple FWI studies have highlighted the regularization benefits of deep reparameterization, the underlying theoretical mechanisms remain insufficiently investigated. This work bridges this knowledge gap through systematic spectral dynamics analysis, demonstrating that deep reparameterization implicitly imposes progressive frequency regularization via the spectral learning properties of deep neural networks. To quantitatively characterize the frequency evolution process, we define the logarithmic power spectrum of velocity models as
    \begin{equation}
        \textbf{m}_{\mathrm{Spec}} = 20\log_{10}(\vert FFT(\textbf{m}) \vert),
    \end{equation}
    where $FFT: \mathbb{R}^{H \times W} \rightarrow \mathbb{C}^{H \times W}$ denotes the 2D Fourier transform operator applied to velocity model $\textbf{m}$. The high-frequency energy ratio metric is formulated as:
    \begin{equation}
        \mathrm{HF\text{-}Ratio}(\textbf{m}) = \frac{\sum_{r>r_c}|FFT(\textbf{m})|}{\sum_{r\geq0}|FFT(\textbf{m})|},
    \end{equation}
    where $r = \sqrt{(u - H/2)^2 + (v - W/2)^2}$ is the radial distance from spectral coordinates to the center, and $r_c = 2.375$ serving as the predefined threshold distinguishing low- and high-frequency components. 
    
\begin{figure}[!ht]
        \centering
        \includegraphics[width=0.5\textwidth]{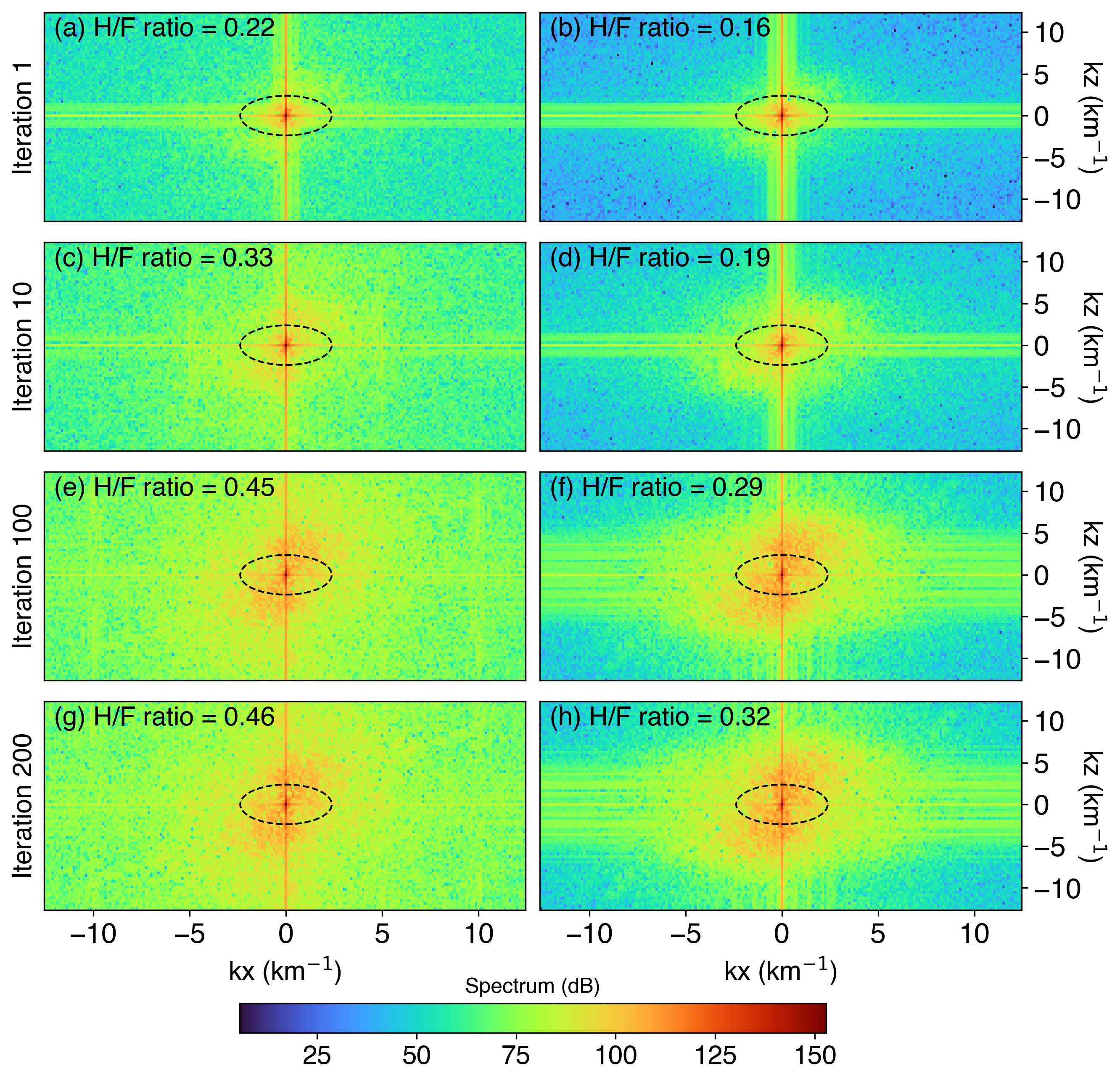}
        \caption{Spectral analysis of traditional FWI and DR-FWI using the CNN-$v_p$ initialization strategy. The left column displays the spectra of inverted velocity models obtained by traditional FWI after (a) 1, (c) 10, (e) 100, and (g) 200 iterations. The right column shows the corresponding spectra from DR-FWI after (b) 1, (d) 10, (f) 100, and (h) 200 iterations. The dashed line at the center of each panel indicates the \(\mathrm{HF\text{-}Ratio}\) boundary, separating low (inner) and high (outer) frequency components.}
        \label{fig:spectram_analysis}
    \end{figure}

   The spectral evolution results are visualized in Fig.~\ref{fig:spectram_analysis}. Conventional FWI (Figs.~\ref{fig:spectram_analysis}a, c, e, g) exhibits simultaneously updates across low and high frequencies, with a sudden injection of high-frequency energy during early iterations (1-10), reflected in an abrupt 10\% surge in \(\mathrm{HF\text{-}Ratio}\). This impulsive spectral perturbation tends to amplify high-frequency noise and drive solutions toward local minima. In contrast, deep reparameterized FWI (Figs.~\ref{fig:spectram_analysis}b, d, f and h) exhibits a staged spectral learning pattern. During the initial low-frequency dominance phase (such as iterations 1-10), spectral energy concentrates in low-frequency band ($r < r_c$); In the subsequent bandwidth expansion phase (such as iterations 100-200), high-frequency components gradually emerge and propagate outward in concentric spectral patterns. This progressive frequency activation aligns with the known low-to-high frequency learning bias of DNN parameterization. Through spectral analysis grounded in DNNs' inherent low-frequency bias, we substantiate that deep reparameterization effectively mitigates cycle-skipping issues by establishing a gradual frequency learning paradigm, contrasting sharply with conventional FWI's instability-prone abrupt frequency injection.

\subsection{Structure Ablation of Multiparameter Reparameterization}

    The incorporation of branch network structures and normalization/denormalization operations extends the deep reparameterization approach to multiparameter FWI. This method enables parameter-specific feature extraction and scale alignment across different physical properties, thereby effectively addressing the crosstalk problem in multiparameter inversion. Its effectiveness has been validated on both the synthetic anomaly and Marmousi2 models. In this section, we investigate the impact of CNN hyperparameters—such as the number of layers in the backbone and branch networks and the number of feature extraction channels—on inversion performance. Two network architectures are compared: (1) a unified backbone-branch CNN that jointly generates all physical parameters (e.g. $v_p$, $v_s$, $\rho$), and (2) multiple independent CNNs, each responsible for generating a single parameter. We perform a systematic search across 24 backbone-branch network configurations. The backbone network consists of 2- and 3-layer CNNs with 64, 128, 256, and 512 channels. The branch networks include 1-, 2-, and 3-layer CNNs with channel settings of \{1\}, \{32, 1\}, and \{64, 32, 1\}, respectively. For comparison, we also evaluate eight independent CNN configurations (multiple independent CNNs) comprising 2-layer and 3-layer CNNs with 32, 64, 128, and 256 channels. 

\begin{figure}[!ht]
        \centering
        \includegraphics[width=0.5\textwidth]{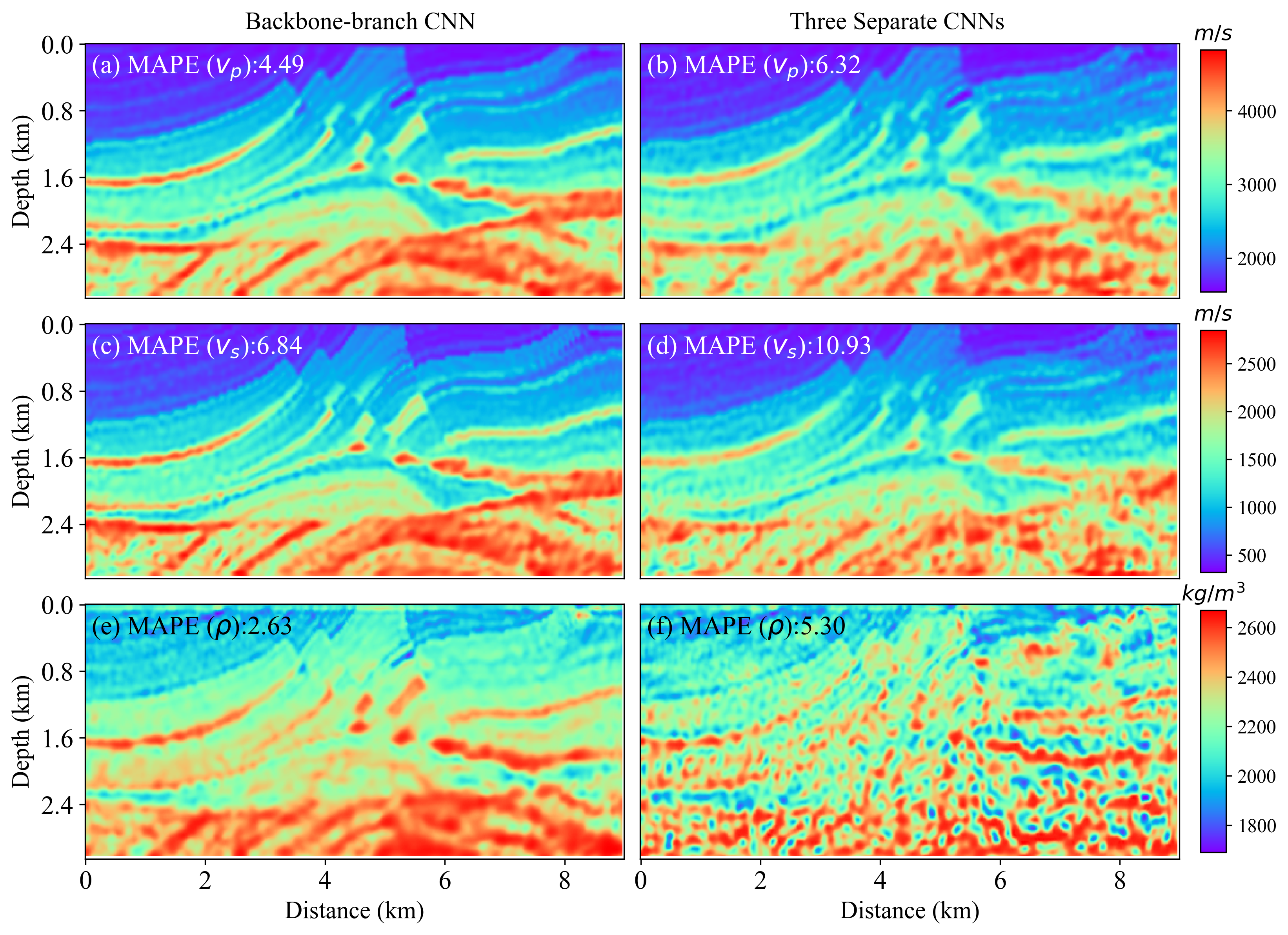}
        \caption{Comparison of inversion results for $v_p$, $v_s$, and $\rho$ obtained from two reparameterization strategies in multiparameter FWI: a single "backbone-branch" network (a, c, e) and three independent CNN networks (b, d, f). Subfigures (a, b) correspond to $v_p$, (c, d) to $v_s$, and (e, f) to $\rho$.}
        \label{fig:MultiParameter_strategy_model}
    \end{figure}

\begin{figure}[!ht]
        \centering
        \includegraphics[width=0.5\textwidth]{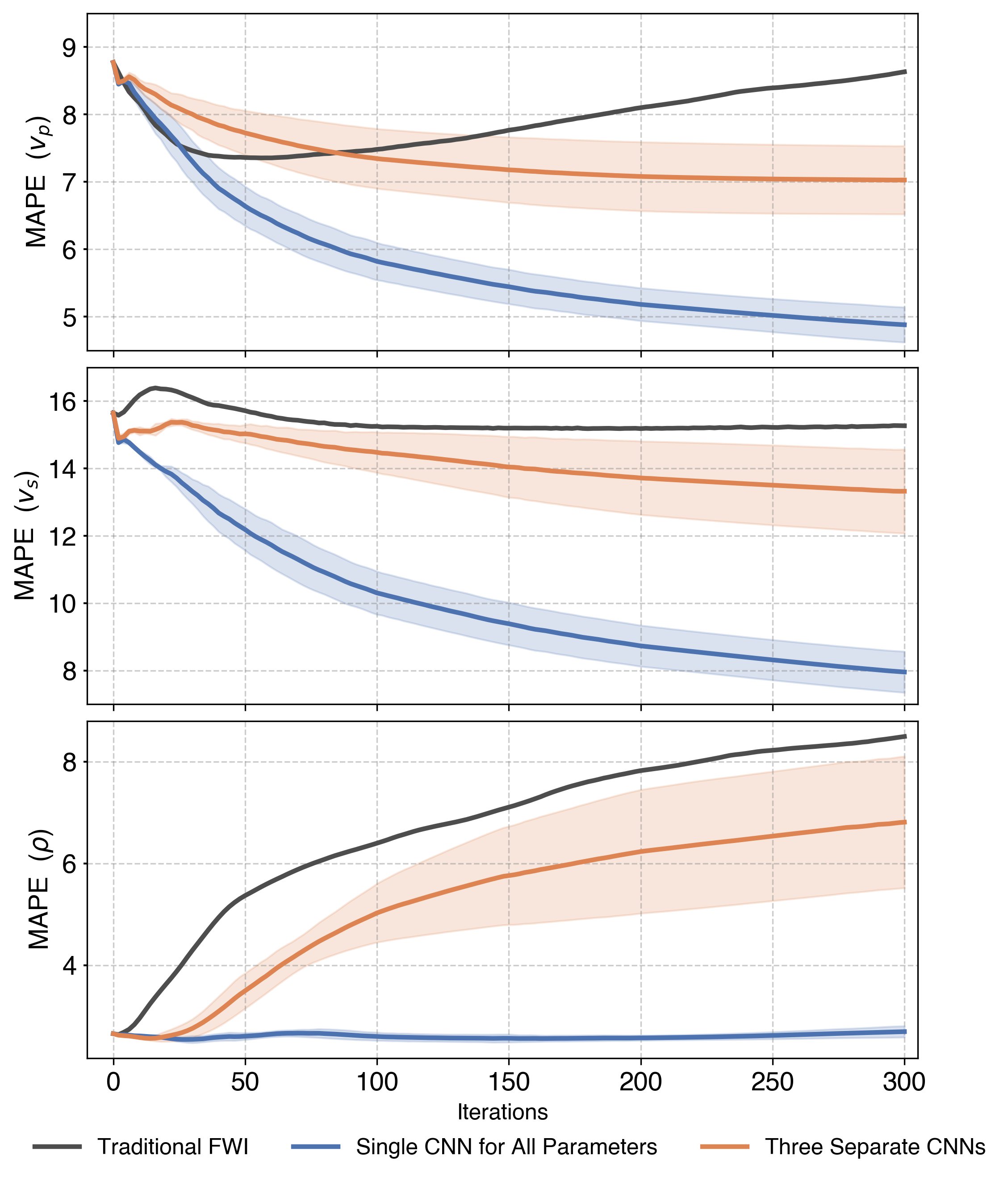}
        \caption{Comparison of MAPE evolution during inversion for (a) $v_p$, (b) $v_s$, and (c) $\rho$ using two reparameterization strategies in multiparameter FWI. Solid lines indicate the mean MAPE across different network configurations, and shaded regions represent the variability of inversion performance under each strategy.}
        \label{fig:MultiParameter_strategy_misfit}
    \end{figure}

    Figs.~\ref{fig:MultiParameter_strategy_model}a, c and e display the best inversion results for $v_p$, $v_s$, and $\rho$ using the backbone-branch reparameterization strategy, while Figs.~\ref{fig:MultiParameter_strategy_model}b, d and f show the best inversion results using three independent CNNs. Fig.~\ref{fig:MultiParameter_strategy_misfit} illustrates the changes of MAPE during the iterative process for different parameterization strategies, where the solid lines represent the mean inversion results for each strategy with varying network configurations, and the shaded areas indicate the distribution of inversion results for different network structures. The comparative results demonstrate that both reparameterization strategies significantly outperform traditional FWI methods. Among them, the "backbone-branch" network exhibits the best performance in recovering different physical parameters, especially in terms of density constraints, where the effect is most pronounced.

\subsection{Limitations and Future Work}
    Extensive ablation studies conducted in this work demonstrate that DR-FWI significantly improves both the accuracy and stability of inversion results. Therefore, DR-FWI can be regarded as a general strategy for full waveform inversion, comparable to widely adopted techniques such as total variation (TV) regularization, multiscale frequency continuation, and gradient post-processing. Prior studies have shown that deep reparameterization networks are highly flexible and can be integrated into both conventional adjoint-state-method-based FWI frameworks \cite{he_2021_Reparameterized, dhara_2022_ElasticAdjointNet} and automatic-differentiation-based deep learning frameworks \cite{zhu_2021_Integrating, liu_2025_Automatic}. Owing to its high compatibility with various FWI paradigms, DR-FWI exhibits strong scalability and plug-and-play capability, facilitating seamless integration into existing inversion workflows with minimal modification.

    Nonetheless, several practical limitations should be acknowledged. When incorporated into traditional FWI frameworks, the optimization of the reparameterization network typically requires GPU acceleration, whereas conventional FWI implementations predominantly rely on multi-core CPU clusters. The resulting data exchange between heterogeneous computing architectures may introduce additional computational overhead. In contrast, integrating DR-FWI into AD-based frameworks allows for unified GPU computation, which is more efficient. However, the substantial memory consumption intrinsic to ADFWI remains a significant bottleneck, especially when applied to large-scale 3-D modeling or field data.

    In terms of applicability, this study has extended the use of DR-FWI beyond its original focus on noise robustness, demonstrating its effectiveness under more challenging conditions, including sparse acquisition and multiparameter inversion. Future efforts will focus on applying DR-FWI to more complex and realistic scenarios, such as low-frequency-missing field data \cite{li_2016_Fullwaveform} and large-scale 3D FWI tasks \cite{ben-hadj-ali_2008_Velocity}. More broadly, deep reparameterization strategies hold promise for a wide range of inverse problems, including surface wave inversion, gravity and magnetic inversion, medical imaging, and atmospheric parameter estimation. Furthermore, the demonstrated capability of DR-FWI to suppress crosstalk in multiparameter inversion opens new avenues for research in multiparameter and multiphysics joint inversion, providing valuable guidance for future theoretical and algorithmic developments.

%
%
\section{CONCLUSION}
    This study systematically evaluates various deep reparameterization network architectures, including U-Net, CNNs, and MLP, as well as different initial model embedding schemes (pretraining-based and direct summation), in the context of full waveform inversion. The results demonstrate that deep reparameterization methods serve as effective plug-and-play modules, significantly enhancing inversion accuracy and stability. Extensive ablation experiments indicate that a multi-layer CNN architecture combined with a pretraining-based initial model embedding strategy yields optimal inversion performance. These findings provide important guidelines for the design of deep reparameterization networks in practical geophysical applications. Moreover, the study investigates the underlying mechanism of DR-FWI's performance improvement by analyzing the the spectral learning behavior of deep neural networks. It is found that during the early stages of inversion, the network tends to capture low-frequency statistical features of the velocity models, and progressively reconstructs high-frequency details as the inversion progresses. This progressive spectral learning not only effectively suppresses noise and refines inversion results, but also provides theoretical support for the robustness of DR-FWI and enhances the interpretability of deep neural network-based inversion methods. From an application perspective, DR-FWI has successfully extended from its initial focus on noise robustness to more challenging scenarios, such as sparse observation and multiparameter inversion. In particular, DR-FWI's strong robustness to handle sparse and noisy data significantly reduces acquisition requirements, making it highly applicable to real-world field conditions. Furthermore, its capability to suppress crosstalk in multiparameter inversion suggests a promising direction for future research in multiparameter and multiphysics joint inversion, offering valuable theoretical insights and architectural foundations for the development of next-generation inversion algorithms.

%
%

\bibliographystyle{IEEEtran}

\bibliography{reference}

\begin{thebibliography}{10}
\providecommand{\url}[1]{#1}
\csname url@samestyle\endcsname
\providecommand{\newblock}{\relax}
\providecommand{\bibinfo}[2]{#2}
\providecommand{\BIBentrySTDinterwordspacing}{\spaceskip=0pt\relax}
\providecommand{\BIBentryALTinterwordstretchfactor}{4}
\providecommand{\BIBentryALTinterwordspacing}{\spaceskip=\fontdimen2\font plus
\BIBentryALTinterwordstretchfactor\fontdimen3\font minus \fontdimen4\font\relax}
\providecommand{\BIBforeignlanguage}[2]{{%
\expandafter\ifx\csname l@#1\endcsname\relax
\typeout{** WARNING: IEEEtran.bst: No hyphenation pattern has been}%
\typeout{** loaded for the language `#1'. Using the pattern for}%
\typeout{** the default language instead.}%
\else
\language=\csname l@#1\endcsname
\fi
#2}}
\providecommand{\BIBdecl}{\relax}
\BIBdecl

\bibitem{lailly_1983_Seismic}
P.~Lailly, ``The seismic inverse problem as a sequence of before stack migrations,'' in \emph{Conference on Inverse Scattering, Theory and Application, Society for Industrial and Applied Mathematics, Expanded Abstracts}, Philadelphia, 1983, pp. 206--220.

\bibitem{tarantola_1984_Linearized}
A.~Tarantola, ``Linearized inversion of seismic reflection data,'' \emph{Geophysical Prospecting}, vol.~32, no.~6, pp. 998--1015, 1984.

\bibitem{virieux_2009_Overview}
J.~Virieux and S.~Operto, ``An overview of full-waveform inversion in exploration geophysics,'' \emph{Geophysics}, vol.~74, no.~6, pp. WCC1--WCC26, 2009.

\bibitem{tromp_2019_Seismic}
J.~Tromp, ``Seismic wavefield imaging of earth's interior across scales,'' \emph{Nature Reviews Earth and Environment}, vol.~1, no.~1, pp. 40--53, 2019.

\bibitem{kalita_2019_Regularized}
M.~Kalita, V.~Kazei, Y.~Choi, and T.~Alkhalifah, ``Regularized full-waveform inversion with automated salt flooding,'' \emph{Geophysics}, vol.~84, no.~4, pp. R569--R582, 2019.

\bibitem{brossier_2009_Seismic}
R.~Brossier, S.~Operto, and V.~Jean, ``Seismic imaging of complex structures by 2d elastic frequency-domain full-waveform inversion,'' \emph{Geophysics}, vol.~74, no.~6, 2009.

\bibitem{routh_2017_Impact}
P.~Routh, R.~Neelamani, R.~Lu, S.~Lazaratos, H.~Braaksma, S.~Hughes, R.~Saltzer, J.~Stewart, K.~Naidu, H.~Averill, V.~Gottumukkula, P.~Homonko, J.~Reilly, and D.~Leslie, ``Impact of high-resolution fwi in the western black sea: Revealing overburden and reservoir complexity,'' \emph{The Leading Edge}, vol.~36, no.~1, pp. 60--66, 2017.

\bibitem{schmelzbach_2016_Advanced}
C.~Schmelzbach, S.~Greenhalgh, F.~Reiser, J.-F. Girard, F.~Bretaudeau, L.~Capar, and A.~Bitri, ``Advanced seismic processing/imaging techniques and their potential for geothermal exploration,'' \emph{Interpretation}, vol.~4, no.~4, pp. SR1--SR18, 2016.

\bibitem{gauthier_1986_Twodimensional}
O.~Gauthier, J.~Virieux, and A.~Tarantola, ``Two-dimensional nonlinear inversion of seismic waveforms: Numerical results,'' \emph{Geophysics}, vol.~51, no.~7, pp. 1387--1403, 1986.

\bibitem{luo_1991_Waveequation}
Y.~Luo and G.~T. Schuster, ``Wave-equation traveltime inversion,'' \emph{Geophysics}, vol.~56, no.~5, pp. 645--653, 1991.

\bibitem{métivier_2014_Full}
L.~M{\'e}tivier, F.~Bretaudeau, R.~Brossier, S.~Operto, and J.~Virieux, ``Full waveform inversion and the truncated newton method: Quantitative imaging of complex subsurface structures,'' \emph{Geophysical Prospecting}, vol.~62, no.~6, pp. 1353--1375, 2014.

\bibitem{tarantola_2005_Inverse}
A.~Tarantola, \emph{Inverse Problem Theory and Methods for Model Parameter Estimation}.\hskip 1em plus 0.5em minus 0.4em\relax {Society for Industrial and Applied Mathematics}, 2005.

\bibitem{operto_2013_Guided}
S.~Operto, Y.~Gholami, V.~Prieux, A.~Ribodetti, R.~Brossier, L.~Metivier, and J.~Virieux, ``A guided tour of multiparameter full-waveform inversion with multicomponent data: From theory to practice,'' \emph{The Leading Edge}, vol.~32, no.~9, pp. 1040--1054, 2013.

\bibitem{yang_2014_Multiparameter}
J.~Yang, Y.~Liu, and L.~Dong, ``A multi-parameter full waveform inversion strategy for acoustic media with variable density,'' \emph{Chinese Journal of Geophysics (in Chinese)}, vol.~57, no.~2, pp. 628--643, 2014.

\bibitem{yu_2021_Deep}
S.~Yu and J.~Ma, ``Deep learning for geophysics: Current and future trends,'' \emph{Reviews of Geophysics}, vol.~59, no.~3, p. e2021RG000742, 2021.

\bibitem{lin_2023_Physicsguided}
Y.~Lin, J.~Theiler, and B.~Wohlberg, ``Physics-guided data-driven seismic inversion: Recent progress and future opportunities in full-waveform inversion,'' \emph{IEEE Signal Processing Magazine}, vol.~40, no.~1, pp. 115--133, 2023.

\bibitem{wu_2020_InversionNet}
Y.~Wu and Y.~Lin, ``Inversionnet: An efficient and accurate data-driven full waveform inversion,'' \emph{IEEE Transactions on Computational Imaging}, vol.~6, pp. 419--433, 2020.

\bibitem{zhang_2020_Datadriven}
Z.~Zhang and Y.~Lin, ``Data-driven seismic waveform inversion: A study on the robustness and generalization,'' \emph{IEEE Transactions on Geoscience and Remote Sensing}, vol.~58, no.~10, pp. 6900--6913, 2020.

\bibitem{rasht‐behesht_2022_Physicsinformed}
M.~Rasht-Behesht, C.~Huber, K.~Shukla, and G.~E. Karniadakis, ``Physics-informed neural networks (pinns) for wave propagation and full waveform inversions,'' \emph{Journal of Geophysical Research: Solid Earth}, vol. 127, no.~5, 2022.

\bibitem{ren_2020_Physicsbased}
Y.~Ren, X.~Xu, S.~Yang, L.~Nie, and Y.~Chen, ``A physics-based neural-network way to perform seismic full waveform inversion,'' \emph{IEEE Access}, vol.~8, pp. 112\,266--112\,277, 2020.

\bibitem{he_2021_Reparameterized}
Q.~He and Y.~Wang, ``Reparameterized full-waveform inversion using deep neural networks,'' \emph{Geophysics}, vol.~86, no.~1, pp. V1--V13, 2021.

\bibitem{zhu_2021_Integrating}
W.~Zhu, K.~Xu, E.~Darve, B.~Biondi, and G.~C. Beroza, ``Integrating deep neural networks with full-waveform inversion: Reparameterization, regularization, and uncertainty quantification,'' \emph{Geophysics}, vol.~87, no.~1, pp. R93--R109, 2021.

\bibitem{sun_2023_Implicit}
J.~Sun, K.~Innanen, T.~Zhang, and D.~Trad, ``Implicit seismic full waveform inversion with deep neural representation,'' \emph{Journal of Geophysical Research: Solid Earth}, vol. 128, no.~3, p. e2022JB025964, 2023.

\bibitem{liu_2025_Automatic}
F.~Liu, H.~Li, G.~Zou, and J.~Li, ``Automatic differentiation-based full waveform inversion with flexible workflows,'' \emph{Journal of Geophysical Research: Machine Learning and Computation}, vol.~2, no.~1, p. e2024JH000542, 2025.

\bibitem{wu_2019_Parametric}
Y.~Wu and G.~A. McMechan, ``Parametric convolutional neural network-domain full-waveform inversion,'' \emph{Geophysics}, vol.~84, no.~6, pp. R881--R896, 2019.

\bibitem{sun_2021_Physicsguided}
J.~Sun, K.~A. Innanen, and C.~Huang, ``Physics-guided deep learning for seismic inversion with hybrid training and uncertainty analysis,'' \emph{Geophysics}, vol.~86, no.~3, pp. R303--R317, 2021.

\bibitem{wang_2023_Prior}
F.~Wang, X.~Huang, and T.~A. Alkhalifah, ``A prior regularized full waveform inversion using generative diffusion models,'' \emph{IEEE Transactions on Geoscience and Remote Sensing}, vol.~61, pp. 1--11, 2023.

\bibitem{taufik_2024_Learned}
M.~H. Taufik, F.~Wang, and T.~Alkhalifah, ``Learned regularizations for multi-parameter elastic full waveform inversion using diffusion models,'' \emph{Journal of Geophysical Research: Machine Learning and Computation}, vol.~1, no.~1, p. e2024JH000125, 2024.

\bibitem{dhara_2022_ElasticAdjointNet}
A.~Dhara and M.~Sen, ``Elastic-adjointnet: A physics-guided deep autoencoder to overcome crosstalk effects in multiparameter full-waveform inversion,'' in \emph{Second International Meeting for Applied Geoscience \& Energy}.\hskip 1em plus 0.5em minus 0.4em\relax Houston, Texas: {Society of Exploration Geophysicists and American Association of Petroleum Geologists}, 2022, pp. 882--886.

\bibitem{alford_1974_ACCURACY}
R.~M. Alford, K.~R. Kelly, and D.~M. Boore, ``Accuracy of finite-difference modeling of the acoustic wave equation,'' \emph{Geophysics}, vol.~39, no.~6, pp. 834--842, 1974.

\bibitem{schuster_2017_Seismic}
G.~T. Schuster, \emph{Seismic Inversion}.\hskip 1em plus 0.5em minus 0.4em\relax Society of Exploration Geophysicists, 2017.

\bibitem{levander_1988_Fourthorder}
A.~R. Levander, ``Fourth-order finite-difference {\emph{p-sv}} seismograms,'' \emph{Geophysics}, vol.~53, no.~11, pp. 1425--1436, 1988.

\bibitem{virieux_1986_PSV}
J.~Virieux, ``{\emph{P-SV}} wave propagation in heterogeneous media: Velocity-stress finite-difference method,'' \emph{Geophysics}, vol.~51, no.~4, pp. 889--901, 1986.

\bibitem{liu_2024_Liufeng2317}
F.~Liu, ``Liufeng2317/adfwi: Zenodo,'' Zenodo, 2024.

\bibitem{bozdağ_2011_Misfit}
E.~Bozda{\u g}, J.~Trampert, and J.~Tromp, ``Misfit functions for full waveform inversion based on instantaneous phase and envelope measurements: Misfit functions for full waveform inversion,'' \emph{Geophysical Journal International}, vol. 185, no.~2, pp. 845--870, 2011.

\bibitem{choi_2012_Application}
Y.~Choi and T.~Alkhalifah, ``Application of multi-source waveform inversion to marine streamer data using the global correlation norm,'' \emph{Geophysical Prospecting}, vol.~60, no.~4, pp. 748--758, 2012.

\bibitem{fichtner_2006_Adjoint}
A.~Fichtner, H.-P. Bunge, and H.~Igel, ``The adjoint method in seismology,'' \emph{Physics of the Earth and Planetary Interiors}, vol. 157, no. 1-2, pp. 86--104, 2006.

\bibitem{liu_2006_FiniteFrequency}
Q.~Liu and J.~Tromp, ``Finite-frequency kernels based on adjoint methods,'' \emph{Bulletin of the Seismological Society of America}, vol.~96, no.~6, pp. 2383--2397, 2006.

\bibitem{sambridge_2007_Automatic}
M.~Sambridge, P.~Rickwood, N.~Rawlinson, and S.~Sommacal, ``Automatic differentiation in geophysical inverse problems,'' \emph{Geophysical Journal International}, vol. 170, no.~1, pp. 1--8, 2007.

\bibitem{herrmann_2023_Use}
L.~Herrmann, T.~B{\"u}rchner, F.~Dietrich, and S.~Kollmannsberger, ``On the use of neural networks for full waveform inversion,'' \emph{Computer Methods in Applied Mechanics and Engineering}, vol. 415, p. 116278, 2023.

\bibitem{lempitsky_2018_Deep}
V.~Lempitsky, A.~Vedaldi, and D.~Ulyanov, ``Deep image prior,'' in \emph{2018 IEEE/CVF Conference on Computer Vision and Pattern Recognition}, 2018, pp. 9446--9454.

\bibitem{sun_2023_FullWaveform}
P.~Sun, F.~Yang, H.~Liang, and J.~Ma, ``Full-waveform inversion using a learned regularization,'' \emph{IEEE Transactions on Geoscience and Remote Sensing}, vol.~61, pp. 1--15, 2023.

\bibitem{zhao_2020_Referencedriven}
D.~Zhao, F.~Zhao, and Y.~Gan, ``Reference-driven compressed sensing mr image reconstruction using deep convolutional neural networks without pre-training,'' \emph{Sensors}, vol.~20, no.~1, p. 308, 2020.

\bibitem{martin_2006_Marmousi2}
G.~S. Martin, R.~Wiley, and K.~J. Marfurt, ``Marmousi2: An elastic upgrade for marmousi,'' \emph{The Leading Edge}, vol.~25, no.~2, pp. 156--166, 2006.

\bibitem{aminsadeh_1996_3D}
F.~Aminsadeh, \emph{3-D Salt and Overthrust Seismic Models}, P.~Weimer and T.~L. Davis, Eds.\hskip 1em plus 0.5em minus 0.4em\relax American Association of Petroleum Geologists, 1996.

\bibitem{gray_1995_Migration}
S.~H. Gray and K.~J. Marfurt, ``Migration from topography: Improving the near-surface image,'' \emph{Canadian Society of Exploration Geophysicists}, vol.~31, pp. 18--24, 1995.

\bibitem{kingma_2017_Adam}
D.~P. Kingma and J.~Ba, ``Adam: A method for stochastic optimization,'' 2017.

\bibitem{hyndman_2006_Another}
R.~J. Hyndman and A.~B. Koehler, ``Another look at measures of forecast accuracy,'' \emph{International Journal of Forecasting}, vol.~22, no.~4, pp. 679--688, 2006.

\bibitem{wang_2004_Image}
Z.~Wang, A.~C. Bovik, H.~R. Sheikh, and E.~P. Simoncelli, ``Image quality assessment: From error visibility to structural similarity,'' \emph{IEEE Transactions on Image Processing}, vol.~13, no.~4, pp. 600--612, 2004.

\bibitem{rahaman_2018_Spectral}
N.~Rahaman, A.~Baratin, D.~Arpit, F.~Draxler, M.~Lin, F.~A. Hamprecht, Y.~Bengio, and A.~Courville, ``On the spectral bias of neural networks,'' 2018.

\bibitem{chakrabarty_2019_Spectral}
P.~Chakrabarty and S.~Maji, ``The spectral bias of the deep image prior,'' 2019.

\bibitem{dokter_2017_Full}
E.~Dokter, D.~K{\"o}hn, D.~Wilken, D.~De~Nil, and W.~Rabbel, ``Full waveform inversion of sh- and love-wave data in near-surface prospecting,'' \emph{Geophysical Prospecting}, vol.~65, no.~S1, pp. 216--236, 2017.

\bibitem{li_2016_Fullwaveform}
Y.~E. Li and L.~Demanet, ``Full-waveform inversion with extrapolated low-frequency data,'' \emph{Geophysics}, vol.~81, no.~6, pp. R339--R348, 2016.

\bibitem{ben-hadj-ali_2008_Velocity}
H.~{Ben-Hadj-Ali}, S.~Operto, and J.~Virieux, ``Velocity model building by 3d frequency-domain, full-waveform inversion of wide-aperture seismic data,'' \emph{Geophysics}, vol.~73, no.~5, pp. VE101--VE117, 2008.

\end{thebibliography}

\vfill
\end{document}